\documentclass[conference]{IEEEtran}
\pdfoutput=1
\usepackage{fancyhdr}

\usepackage{todonotes}
\usepackage{esvect}
\usepackage{graphicx}
\usepackage{textcomp}
\usepackage{tabularx}
\usepackage{xcolor}
\usepackage{mwe}
\usepackage{subcaption}
\usepackage{caption}
\usepackage{xcolor}
\usepackage{algorithm}
\usepackage{algpseudocode}
\usepackage{xspace}
\usepackage{paralist}
\usepackage{booktabs}
\usepackage{multirow}
\usepackage{cprotect}
\usepackage{comment}
\usepackage{multicol}
\usepackage{soul}
\usepackage{listings}
\usepackage{listing}
\usepackage{subcaption}
\usepackage[nounderscore]{syntax}
\usepackage{hyperref}

\usepackage{cite}
\usepackage{amsmath,amssymb,amsfonts}
\usepackage{graphicx}
\usepackage{textcomp}
\usepackage{xcolor}
\usepackage{booktabs}
\usepackage{tagging}
\usepackage{paralist}

\usepackage{sc24repro}
\usepackage{hyperref}

\AtBeginDocument{\DeclareCaptionSubType{lstlisting}}

\newcolumntype{P}[1]{>{\RaggedRight\hspace{0pt}}p{#1}}
\def\BibTeX{{\rm B\kern-.05em{\sc i\kern-.025em b}\kern-.08em
    T\kern-.1667em\lower.7ex\hbox{E}\kern-.125emX}}
    
\definecolor{belizehole}{HTML}{2980b9}
\definecolor{clouds}{HTML}{f4f8f9}
\definecolor{midnightblue}{HTML}{2c3e50}
\definecolor{pomegranate}{HTML}{c0392b}
\definecolor{royalblue}{HTML}{3867d6}
\definecolor{greensea}{HTML}{16a085}
\definecolor{nephritis}{HTML}{27ae60}
\definecolor{amethyst}{HTML}{9b59b6}

\definecolor{mGreen}{rgb}{0,0.6,0}
\definecolor{mGray}{rgb}{0.5,0.5,0.5}
\definecolor{mPurple}{rgb}{0.58,0,0.82}
\definecolor{backgroundColour}{rgb}{0.95,0.95,0.92}
\definecolor{calpolypomonagreen}{rgb}{0.12, 0.3, 0.17}
\definecolor{cadmiumgreen}{rgb}{0.0, 0.42, 0.24}
\definecolor{light-gray}{gray}{0.9}

\lstdefinestyle{SyntaxStyle}{
	basicstyle=\footnotesize,
	escapeinside={<@}{@>}, 
    commentstyle=\color{green},
    keywordstyle=\color{blue},
    numberstyle=\tiny\color{gray},
    stringstyle=\color{mPurple},
    breakatwhitespace=false,         
    breaklines=true,                 
    captionpos=b,                    
    keepspaces=true,                 
    numbers=none,                    
    numbersep=5pt,                  
    showspaces=false,                
    showstringspaces=false,
    showtabs=false,                  
    tabsize=2,
    language=C,
    morekeywords={size_t, int8_t},
    xleftmargin=1em,
    showlines=true
}

\lstdefinestyle{CStyle}{
	basicstyle=\footnotesize,
	escapeinside={<@}{@>}, 
    commentstyle=\color{green},
    keywordstyle=\color{blue},
    numberstyle=\tiny\color{gray},
    stringstyle=\color{mPurple},
    breakatwhitespace=false,         
    breaklines=true,                 
    captionpos=b,                    
    keepspaces=true,                 
    numbers=left,                    
    numbersep=5pt,                  
    showspaces=false,                
    showstringspaces=false,
    showtabs=false,                  
    tabsize=2,
    language=C,
    morekeywords={size_t, int8_t},
    xleftmargin=1em,
    showlines=true
}
\definecolor{celadon}{rgb}{0.67, 0.88, 0.69}
\definecolor{amber}{rgb}{1.0, 0.49, 0.0}
\definecolor{bananamania}{rgb}{0.98, 0.91, 0.71}
\definecolor{beaublue}{rgb}{0.74, 0.83, 0.9}
\definecolor{bubblegum}{rgb}{0.99, 0.76, 0.8}
\definecolor{lightgray}{rgb}{0.83, 0.83, 0.83}

\usepackage{ifthen} %
\usepackage{xcolor} %
\newboolean{colorText}
\setboolean{colorText}{true} %
\newcommand{\coloredtext}[2]{%
  \ifthenelse{\boolean{colorText}}%
  {\textcolor{#1}{#2}}%
  {#2}%
}

\usepackage{subcaption}

\usepackage[frozencache,cachedir=.,outputdir=.]{minted}
\usepackage{enumitem}

\def\hpacml{\textsc{hpac-ml}\xspace}

\hypersetup{
    colorlinks,
    linkcolor={red!50!black},
    citecolor={blue!50!black},
    urlcolor={blue!80!black}
}

\newif\ifincludeappendix
\includeappendixtrue

\begin{document}
\title{HPAC-ML: A Programming Model for Embedding ML Surrogates in Scientific Applications}

\author{
    \IEEEauthorblockN{Zane Fink\IEEEauthorrefmark{1}, Konstantinos Parasyris\IEEEauthorrefmark{2}, Praneet Rathi\IEEEauthorrefmark{1}}
    \IEEEauthorblockN{Giorgis Georgakoudis\IEEEauthorrefmark{2}, Harshitha Menon\IEEEauthorrefmark{2}, Peer-Timo Bremer\IEEEauthorrefmark{2}}
    \IEEEauthorblockA{\IEEEauthorrefmark{1}University of Illinois Urbana-Champaign
    \\\{zanef2, prathi3\}@illinois.edu}
    \IEEEauthorblockA{\IEEEauthorrefmark{2}Lawrence Livermore National Laboratory
    \\\{parasyris1, georgakoudis1, harshitha, bremer5\}@llnl.gov}
}

\maketitle

\thispagestyle{fancy}
\lhead{}
\rhead{}
\chead{}
\lfoot{\footnotesize{
SC24, November 17-22, 2024, Atlanta, Georgia, USA
\newline 979-8-3503-5291-7/24/\$31.00 \copyright 2024 IEEE}}
\rfoot{}
\cfoot{}
\renewcommand{\headrulewidth}{0pt}
\renewcommand{\footrulewidth}{0pt}

\begin{abstract}

Recent advancements in Machine Learning (ML) have substantially improved its predictive and computational abilities, offering promising opportunities for surrogate modeling in scientific applications. By accurately approximating complex functions with low computational cost, ML-based surrogates can accelerate scientific applications by replacing computationally intensive components with faster model inference. However, integrating ML models into these applications remains a significant challenge, hindering the widespread adoption of ML surrogates as an approximation technique in modern scientific computing.

We propose an easy-to-use directive-based programming model that enables developers to seamlessly describe the use of ML models in scientific applications. The runtime support, as instructed by the programming model, performs data assimilation using the original algorithm and can replace the algorithm with model inference. Our evaluation across five benchmarks, testing over 5000 ML models, shows up to 83.6$\times$ speed improvements with minimal accuracy loss (as low as 0.01 RMSE).
\end{abstract}


\IEEEpeerreviewmaketitle

\section{Introduction}

Predictive simulation is a cornerstone of scientific progress in several fields ranging from quantum chemistry to national security, where the speed of simulations directly influences the pace of discovery. 
Until recently, hardware and software innovations drove progress with increased simulation speeds. However, the deceleration of hardware technology advancements, marked by the end of Dennard scaling and the slowing of Moore's Law, poses a significant challenge to maintaining this pace of acceleration. 

Approximate computing --- a promising direction in the post-Moore computing era --- reduces application accuracy for increased performance and energy efficiency. Approximations can be application-specific, such as Newton's method~\cite{Galantai2000:Theory} and polynomial approximations~\cite{Funaro2008:Polynomial}, or general-purpose techniques such as loop perforation~\cite{Sidiroglou2011:Managing,Hoffmann2009:Using}, precision scaling~\cite{Menon2018:ADAPT,Laguna2019:GPUMixer}, and memoization~\cite{mishra2014iact,tziantzioulis2018temporal}. More recently, the widespread adoption of specialized accelerators in HPC and the growing prominence of Machine Learning (ML) has led to the exploration of using Neural Network (NN) based surrogate models to accelerate scientific simulations. One of the key benefits of NN-based approximations is that NN are universal function approximators: with sufficient data and a sufficiently complex network, they can approximate any function. Furthermore, neural networks offer a rich space of architectures such as MLPs, CNNs, and RNNs, providing flexibility and adaptability, thus making them a versatile and efficient option for many scientific applications.
As a result, NN surrogates have achieved wide success in several domains, including radiative transfer calculations~\cite{pal2019using},  computational fluid dynamics~\cite{kochkov2021machine}, and Molecular Dynamics (MD) simulations, demonstrating at least 10$\times$ speedup with low error.

Despite ML's performance benefits, collecting training data and integrating such a model in a scientific application is a manual, complicated, and error-prone process.
Since scientific applications are usually developed in high-performance languages such as C/C++ and FORTRAN, ML models must be invoked within the programming language's ecosystem. This requires application and ML expertise.
For example, transforming data to the tensor format expected by ML models requires intimate knowledge of data transformations, increasing the complexity of model integration in scientific codes.

Therefore, the adoption and efficient use of ML models in scientific applications is slow, hindering the exploration of their applicability to diverse applications and situations. The development complexity of such an ML-enhanced application has gone unchecked. With no additional support to ease the integration of ML models in scientific applications, the adoption rate of ML in HPC is slow, and the HPC community is missing out on potential benefits. Moreover, progress in understanding and overcoming the limitations is hindered by the engineering efforts required to embed models in applications.

We present \hpacml{}, a directive-based programming model that enables easy use of ML models in scientific applications. The programming model provides primitives to invoke a model or collect data during application execution to 
be used during offline training of machine learning models. \hpacml{} provides succinct extensions to describe transformations between application and ML model memory layout. The runtime system can schedule execution on both CPUs and GPUs, transparently handling data transfers.
We extend HPAC~\cite{parasyris2021hpac, fink2023:hpacoffload} and thus rely on the mature LLVM infrastructure. As an inference engine, the runtime system uses Torch (C++ interface to PyTorch). Thus, the proposed approach can digest the latest advancements in compilation and AI/ML technologies.

The programming model facilitates our core mechanism to explore the trade-offs between model size, model architecture, performance benefits, and accuracy degradation in 5 HPC benchmarks. Our evaluation presents how well-defined programming abstractions can help automate the search for appropriate model architectures. Instead of manually exploring these trade-offs, we employ Bayesian Optimization to search the space of possible machine learning models, tracking both performance gains and accuracy loss across all benchmarks. 
Briefly, our contributions are the following:
\begin{itemize}
    \item \hpacml{}, an annotation-based programming model for embedding ML surrogates in scientific applications. Application developers annotate parts of their application to replace them with a surrogate model. \hpacml{} enables developers to describe the data flow between application memory and ML data layouts using
    succinct syntax. 
     This scheme supports supervised machine learning workflows while maintaining modularity and separation of concerns: no single person is required to have both application expertise and experience with ML infrastructure.
    The same source code can collect training data and later load a trained surrogate model that replaces the annotated code region, accelerating the application. 
    \item An implementation of the \hpacml{} programming model in the production-level Clang/LLVM compiler with a runtime system that uses industry-standard infrastructure such as Torch for inference and HDF5 for data collection. The implementation is open-source software available to the community\footnote{\url{https://github.com/LLNL/HPAC}}.
    \item A thorough evaluation on a representative suite of GPU-based HPC mini-apps and benchmarks.  Overall, we find up to $83.6\times$ speedup while maintaining output quality with RMSE as low as $0.001$, while the layout transformations add negligible overhead.
    \item \hpacml{} enables a large-scale exploration of the trade-offs between model size, inference speed, and output quality loss. Across thousands of models tested in our evaluation, we derive insightful observations. For instance, for ParticleFilter 
    we identify models with better accuracy than the existing approximation that yield an end-to-end speedup of $9\times$.
\end{itemize}

\section{Background and Related Work}\label{sec:related}

Approximate computing (AC) increases computing power and energy efficiency. Several AC hardware and software strategies demonstrate its utility. Hardware approximation techniques~\cite{froehlich2018towards, Rehman2016,kulkarni2011trading,kahng2012accuracy, gupta2011impact, chippa2014scalable, esmaeilzadeh2012architecture,hedge1999energy,sampson2011enerj, parasyris2017significance} specialized hardware components or modify underlying hardware to accelerate computations while introducing controlled errors. For instance, approximate adders/multipliers~\cite{froehlich2018towards, Rehman2016,kulkarni2011trading, kahng2012accuracy, gupta2011impact} perform add and multiply operations approximately. 
Software approximation techniques focus on algorithmic and software modifications that introduce approximations to gain performance. Examples include loop perforation~\cite{Hoffmann2009, goiri2015approxhadoop}, which skips certain loop iterations, memoization~\cite{michie1968memo, keramidas2015clumsy, samadi2014paraprox}, which caches previous previous results, and significance based computing~\cite{vassiliadis2016towards,vassiliadis2016exploiting,vassiliadis2015programming} which assign scores to computations in respect to their contribution to the final output quality. 
Additionally, there are several works that use Neural Networks (NNs) to approximate code regions, accelerating computations. Grigorian et al.~\cite{grigorian2015brainiac} proposed BRAINIAC that used both precise and NN-based approximate accelerators to achieve significant performance gains. McAfee and Olukotun proposed EMEURO~\cite{mcafee2015emeuro}, a NN-based emulation and acceleration platform. Moreau et al.~\cite{Moreau2015:SNNAP} presented a technique for neural acceleration on an FPGA.  Esmaeilzadeh et al.~\cite{esmaeilzadeh2012neural} introduced a programming model that enabled users to offload approximable code regions to Neural Processing Units (NPUs). Ho et al.~\cite{Ho2022:Tensorox} proposed Tensorox to exploit the half-precision tensor cores available on GPUs to perform approximate computations. These works have focused primarily on leveraging hardware neural networks as accelerators for approximable programs, and hence their applicability for integrating ML libraries into scientific codes is limited.
Many frameworks integrate AC techniques into codes. These include HPAC~\cite{parasyris2021hpac}, HPAC-Offload~\cite{fink2023:hpacoffload}, ApproxHPVM~\cite{sharif2019approxhpvm}, ACCEPT~\cite{sampson2015accept}, and GREEN~\cite{baek2010green}. 
Only HPAC-Offload and SAGE support AC on GPUs. While these frameworks provide valuable support for several approximate computing techniques, they don't support surrogate models, ultimately hindering the adoption of ML models in HPC applications. 

HPAC allows accurate and approximate versions of the same algorithm to coexist in the same binary by viewing them as different \emph{execution paths}. The path taken is dynamically decided at runtime based on activation criteria for approximation. If the activation criteria are unmet, then the \emph{accurate execution path} is taken, and the original code is executed. Otherwise, the \emph{approximate execution path}, which implements the approximation, is taken.
HPAC provides data-flow semantics\cite{Parasyris2022:Approximate} to the developer that describe which application memory regions are accessed by the accurate execution path.

Recently, Dong et al.~\cite{Dong2023:AutoHPCNet} proposed Auto-HPCnet, which uses a client-server protocol to enable NN-based surrogate models, losing integrated support and incurring communication costs between the client and server. Myers et al.~\cite{Myers2024:Amrex} developed Python bindings as a bridge between AMR simulation data and data science software. They provide a flexible approach for representing simulation data and supporting routines written in Python; however, they don't exploit the latest advances in state-of-the-art ML libraries
 such as PyTorch~\cite{paszke2019pytorch}, TensorFlow~\cite{abadi2016tensorflow}, ONNX~\cite{Bai2019:ONNX}, and LBANN~\cite{VanEssen2015:LBANN}. These ML frameworks allow for a trained model to be encoded in a language-agnostic JIT-able model formats, such as TorchScript JIT. The model can be loaded using C++ or Fortran and invoked to perform inference at application execution time. %

Our framework extends HPAC and provides a unified approach to seamlessly use NN surrogate models within an application.  
The approach separates concerns and can be used as a core mechanism, as done in this work, to build automation and study trade-offs between model accuracy, application performance, and application accuracy.

\section{\hpacml{} Programming Model}

Supervised machine learning consists of three phases: data collection, training, and inference. 
The current practice of using NN models in applications follows multiple manual, error-prone sequential steps. Further, it requires rare interdisciplinary expertise. For example, computer scientists instrument scientific applications for data collection and provide the data to ML experts who build and train NN models. These models are integrated into the application by computer scientists. Domain scientists (e.g., physicists) then evaluate the validity of the ML-enhanced application's results. Simply put, there are too many handshakes in this workflow.

The approach does not scale, limiting scientific throughput. Thus \hpacml{},
 facilitates smooth integration of
ML surrogates in applications by providing primitives to capture  
the developer's intent to accelerate the application via NN model usage. 
\hpacml{}'s design abstracts the engineering support 
required to use ML models in applications and insulates the user from knowing \emph{how} the
software system works. Instead, ML and domain scientists
can focus on building more accurate models and validating the ML-enhanced application results.

\begin{figure*}
   \centering
   \includegraphics[width=0.98\linewidth]{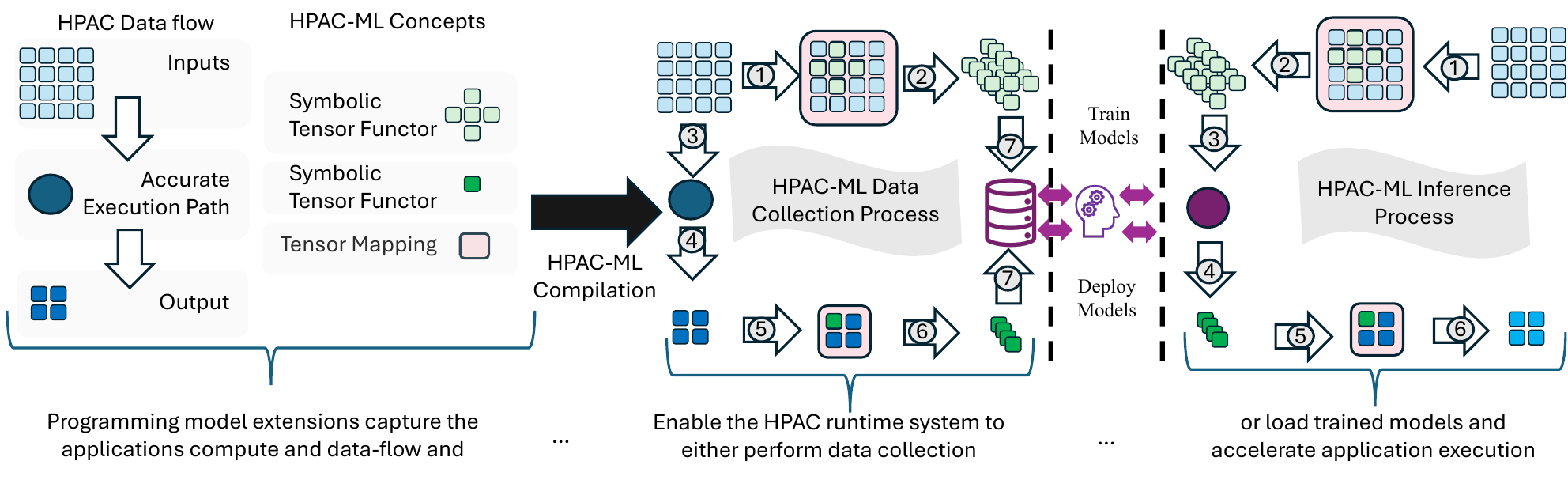}
   \caption{\hpacml{} overview.}
   \label{fig:hpac_ml_overview}
   \vspace{-1em}
\end{figure*}

\subsection{\hpacml{} Design}
The main challenge in abstracting ML use is twofold. 
First, the application developer needs primitives to specify the source code location, often called the code region, to be replaced by the ML model. 
Further primitives are required to define the application data that the code region consumes and produces.
Yet, describing only the coarse data flow of a code region is not sufficient.  The application
data layout and the layout required by modern NN inference engines differ. Thus, a generic
approach needs to provide primitives to the developer to describe the application data layout
and flow at the boundaries of the replaced code regions.

Specifically, from the application code perspective, we need abstractions for two key components: a data bridge layer and an execution control layer. The data bridge layer handles the mismatch between application data and ML frameworks, which expect dense, contiguous inputs and produce dense, contiguous outputs.
The execution control layer is responsible for either performing data collection or model inference. 
The \hpacml{} approach, depicted in Figure~\ref{fig:hpac_ml_overview}, integrates the requirements of both the application and surrogate models in a framework-agnostic way.

\subsubsection{Data Bridge}
The data bridge connects two distinct memory spaces: the application memory space and the tensor memory space, supporting memory access patterns common in scientific applications. For instance, computational fluid dynamics and finite difference methods employ a stencil access pattern, where the value at each gridpoint is updated based on the values of neighboring gridpoints.  Unstructured mesh codes use indirection and pointer-chasing to resolve the map between application data structures and the computational models used in simulations, thereby efficiently bridging the gap between their geometrical representation and the algorithms that operate on them. ML applications such as clustering compare all points in some space to the same set of queries. In such cases, it is not enough to simply wrap application memory arrays in ML library tensors -- there is a difference between how data are accessed by the application and how data are organized in memory. Thus, the data bridge uses two concepts to handle each component separately: a tensor functor and a tensor map.

The tensor functor is an abstraction that lets the user specify how individual data elements are accessed in the application memory space and how these elements should be organized to form a tensor entry in the tensor memory space. Through a process we call memory concretization, the tensor map applies the tensor functor to application memory, completing the bridge between the application and tensor memory spaces and mapping memory between them.
For example, in Figure~\ref{fig:hpac_ml_overview} (left), the developer describes a tensor representing a 
stencil and a tensor mapping. At execution time \hpacml{} (Figure~\ref{fig:hpac_ml_overview} middle, right)  uses both the tensor map and the tensor functor in Steps 1 and 2 to extract tensors representing a 5-point stencil from a contiguous memory grid. The reverse operation transforms the data layout from the model to the application layout (Steps 5 and 6 in Figure~\ref{fig:hpac_ml_overview} middle and right).

\subsubsection{Execution Control}
The execution control layer leverages existing HPAC primitives. 
Briefly, when using HPAC the developer annotates a code region as approximate and specifies the data flow --- the memory locations used as inputs by the annotated region and the memory locations of the computed outputs --- of the code region. With this information, HPAC generates two separate execution paths: an \emph{accurate execution path} with the original application code and the \emph{approximate execution path} implementing the approximation.

\hpacml{} extends the accurate execution path to perform data collection during application execution time as presented in Figure~\ref{fig:hpac_ml_overview} (middle). Before executing 
the accurate execution path, \hpacml{} applies the tensor functors to the region's inputs and maps them into concrete tensors (Steps 1 and 2). These tensors are cached in 
memory. Then, \hpacml{} invokes the accurate execution path, and the application computes 
the output values. Before giving execution control back to the application, \hpacml{} transforms the output memory locations using functors and tensor mapping operations (Steps 5 and 6).
Lastly, \hpacml{} stores this data into a persistent storage database and continues execution.

The persistent storage database can be accessed by the ML expert to train a NN approximation of the accurate execution path. Once the ML experts produce a model of sufficient accuracy they can store it in the filesystem and expose the model path to \hpacml{}. In such a case, as presented in Figure~\ref{fig:hpac_ml_overview} (right), 
during application execution, \hpacml{} reads the model representation and uses the model as the approximate execution path of HPAC. Before invoking the model, \hpacml{} uses developer-defined tensor functors and mapping operations provided by the developer to transform data between the application data layout and model layout.

To summarize, the execution control of \hpacml{} is responsible for either performing data collection or model inference, whereas the data bridge spans the gap between data layouts.

\begin{figure}[t]
\centering
\begin{minted}[fontsize=\footnotesize,linenos, 
escapeinside=@@, numbersep=1pt, 
highlightcolor=celadon!50,highlightlines={1-5}
]{c++}
#pragma approx tensor functor(ifnctr: @\label{ex_functor_decl}@ \
    [i, j,  0:5] = ( ([i-1, j], [i+1, j], \ 
    [i, j-1:j+2]))
#pragma approx tensor functor(ofnctr: @\label{ex_functor_decl_opt}@ \
    [i, j, 0:1] = ([i, j])) 
\end{minted}
\begin{minted}[fontsize=\footnotesize,linenos, 
escapeinside=@@, numbersep=1pt, firstnumber=6,
highlightcolor=bubblegum!50,highlightlines={4-10}
]{c++}
#pragma approx tensor map(to: @\label{ex_map_to_clause_line}@ \
    ifnctr(t[1:N-1, 1:M-1])) 
#pragma approx tensor map(from: @\label{ex_map_from_clause_line}@ \
        ofnctr(tnew[1:N-1, 1:M-1])) 
\end{minted}
\begin{minted}[fontsize=\footnotesize,linenos, 
escapeinside=@@, numbersep=1pt, firstnumber=10,
highlightcolor=beaublue!50,highlightlines={10-11}
]{c++}
#pragma approx ml(predicated:true) in(t) out(tnew) @\label{ex_inference_call}@ \
        db("/path/data.h5") model("/path/model.pt")
    do_timestep(t, tnew);
    
swap(t, tnew);
\end{minted}
\cprotect\caption{\hpacml{} example program. This example replaces the main computation of a 2-D stencil code with NN surrogate inference.
{\sethlcolor{celadon} \hl{{green}}}, highlights the declaration of tensor-functors,
{\sethlcolor{bubblegum} \hl{red}}, highlights the declaration of tensor maps,
{\sethlcolor{beaublue} \hl{blue}}, highlights the declaration of a region from which \hpacml{} will either collect data from or completely replace with a NN model.
}
\label{fig:hpacml_tensor_functor_example_1}
\end{figure}

\subsection{\hpacml{} Example}\label{sub:prog_model}

The \hpacml{} programming model provides concrete realizations of the data bridge and execution control abstractions.
 Figure~\ref{fig:hpacml_tensor_functor_example_1} depicts an example where the directive on Line~\ref{ex_functor_decl} creates a tensor functor named \verb|ifnctr| and applies it to application memory on Line~\ref{ex_map_to_clause_line}, mapping application memory to the tensor memory space. On Line~\ref{ex_map_from_clause_line}, a mapping from the tensor memory space to the application memory space is created. The tensor functor declaration has two main components: a left-hand side (LHS) and a right-hand side (RHS).

The LHS defines an abstract slice that describes a single element access for some symbolic values \verb|i, j|. In Figure~\ref{fig:hpacml_tensor_functor_example_1}, the LHS declares a 3-dimensional tensor with two spatial dimensions and one feature dimension. The user specifies this shape using $i$ and $j$ and the range $0$:$5$. The names $i$ and $j$ are not declared in the program; they are considered symbols for any values $i$ and $j$ may take when the functor is applied to data. The RHS component specifies the origin of each feature without associating the indices with any data. %
Applying a tensor functor to data arrays moves them between the application and tensor memory spaces, completing the data bridge. 
Line~\ref{ex_map_to_clause_line} of Figure~\ref{fig:hpacml_tensor_functor_example_1} does this, where the user maps data \verb|to| the tensor space by applying \verb|functr| to the array \verb|t|.
This is where the range of values are assigned to the symbols of the \verb|ifnctr| functor. Here, $i$ goes from $1$ to $N-1$; $j$ is similarly concretized and mapped to the range $1:M-1$. Simply put, the tensor map performs a sweep over memory, and the functor describes the accesses at each point in the sweep.

Finally, the developer uses the \verb|ml| clause with the \verb|predicated| modifier with the condition set to \verb|true| on Line~\ref{ex_inference_call}. Thus, \hpacml{}
uses the model stored under \verb|"path/model.pt"| and performs NN approximation for every instantiation of the region. 
The \verb|ml| clause accepts a boolean expression evaluated at every invocation of the region. 
The developer can expose this parameter, for example, to the command line interface
and toggle between inference and data collection without requiring recompilation.
For explicit static control of model inference or data collection, the developer can use \verb|infer| or \verb|collect| to enable
inference and data collection for all instantiations of the code region.

\subsection{\hpacml{} Programming Model Grammar}
To further understand the programming model, we show the full syntax of the \verb|tensor functor| and \verb|tensor map| directives along with the \verb|ml| clause in Figure~\ref{lst:syntax}. 

The tensor functor directive (Figure~\ref{lst:syntax} top) is powered by the symbolic slice specifier, or \verb|ss-specifier|, that implements the standard slice syntax with extensions: symbolic constants and expressions that use them. The symbolic constant (\verb|s-constant|) terminal serves as a placeholder within the tensor \verb|ss-specifier|, adapting its value to match the data dimensions provided when the functor is applied to memory. These constants can be used in arithmetic expressions with regular integer expressions to describe different offsets and access patterns involving symbolic constants. For example, in the slice specifier \verb|[i, j, 0:5]| taken from the RHS of Figure~\ref{fig:hpacml_tensor_functor_example_1}, the symbols
\verb|i| and \verb|j| correspond to an \verb|s-constant| whereas \verb|0| and \verb|5| correspond to integer expressions.

The \verb|tensor map| (Figure~\ref{lst:syntax} middle) transforms data from the application layout \verb|to| the ML data layout or
\verb|from| the ML layout to the application as described by the \verb|direction-specifier|. 
The functor application expression (\verb|fa-expr|) applies a declared tensor functor identified by the \verb|decl-functor-id| to a list of user-specified ranges (\verb|map-target-list|). The list contains a sequence of \verb|map-target| expressions.
Every \verb|map-target| initially contains a reference to some array (the array to be transformed by the functor) and a series of concrete-slice-specifiers (\verb|cs-specifier|).
A concrete specifier may only contain expressions that use either integer constants or refer to declared integer variables. The \verb|cs-specifier| expresses ranges of possible values. For example, \verb|[1:N]| describes a range starting from \verb|1| and going to exclusive \verb|N| with a step of \verb|1|.

\begin{figure}
\footnotesize
\begin{minipage}[t]{\linewidth}
\begin{lstlisting}[style=SyntaxStyle]
#pragma approx tensor functor(decl<@-@>functor<@-@>id: 
    ss<@-@>specifier = (ss<@-@>specifier))

ss<@-@>specifier ::= '[' s<@-@>slice, ... ']'
s<@-@>slice ::= s<@-@>expr [ : [s<@-@>expr] [ : [s<@-@>expr]]]
s<@-@>expr ::= s<@-@>expr | s<@-@>constant | <@int@><@-@>expr
\end{lstlisting}
\hrule
\begin{lstlisting}[style=SyntaxStyle]
#pragma approx tensor map(direction<@-@>specifier:
    fa<@-@>expr)

fa<@-@>expr ::= decl<@-@>functor<@-@>id '(' map<@-@>target<@-@>list ')'
map<@-@>target<@-@>list ::= map<@-@>target ...
map<@-@>target ::= array<@-@>ref 
                '[' cs<@-@>specifier | map<@-@>target ']'
cs<@-@>specifier ::= c<@-@>slice [, c<@-@>slice]... 
c<@-@>slice ::= c<@-@>expr [ : [c<@-@>expr] [ : [c<@-@>expr]]]
c<@-@>expr ::= c<@-@>expr | <@int@><@-@>expr
direction<@-@>specifier ::= 'to' | 'from' 
\end{lstlisting}
\hrule
\begin{lstlisting}[style=SyntaxStyle]
#pragma approx ml(ml<@-@>mode [:bool-expr]) [in(mapped<@-@>memory) 
  out(mapped<@-@>memory)] [inout(mapped<@-@>memory)] model(string<@-@>expr) database(string<@-expr@>) if(bool<@-@>expr)
  structured<@-@>block

  mapped<@-@>memory ::= fa<@-@>expr | mapped<@-@>target<@-@>list
  mapped<@-@>target<@-@>list ::= array<@-@>ref [, mapped<@-@>target<@-@>list]
  ml<@-@>mode ::= 'infer' | 'collect' | 'predicated'
\end{lstlisting}
\end{minipage}
\caption{Syntax for the tensor functor declaration (top), tensor mapping (middle), and ml (bottom) clauses of the \hpacml{} programming model.}
\label{lst:syntax}
\end{figure}

The \verb|approx| directive combined with the \verb|ml| clause specify regions of application code to replace with NN surrogate models. 
The syntax of the \verb|approx ml| clause is shown in Figure~\ref{lst:syntax} (bottom). The \verb|ml-mode| is a keyword that accepts \verb|infer,| \verb|collect,| and \verb|predicated|. \verb|infer| instructs \hpacml{} to perform NN inference --- replacing the wrapped structured block with the surrogate model stored at the path provided in the \verb|model| clause. The \verb|collect| clause instructs \hpacml{} to run the structured block and store the inputs and outputs 
in a file under the location specified by the \verb|database| clause.
Finally, the \verb|predicated| clause performs model inference or data collection depending on the boolean expression \verb|ml-cond|.
This enables the dynamic toggling of data collection and model inference.
The \verb|mapped|\verb|-|\verb|target|\verb|-|\verb|list| is a list of references to arrays mapped using the \verb|tensor| \verb|map| clause. The mapped target lists can be designated as input to the NN/structured block (\verb|in|), output from the NN/structured block (\verb|out|), or both (\verb|inout|).

\section{\hpacml{} Implementation}
The \hpacml{} compiler and runtime system implement the data bridge and execution control layers. The \hpacml{} compiler extends the HPAC compiler implemented in Clang/LLVM v17.

\subsection{Data Bridge}

\begin{figure*}
   \centering
   \includegraphics[width=0.88\linewidth]{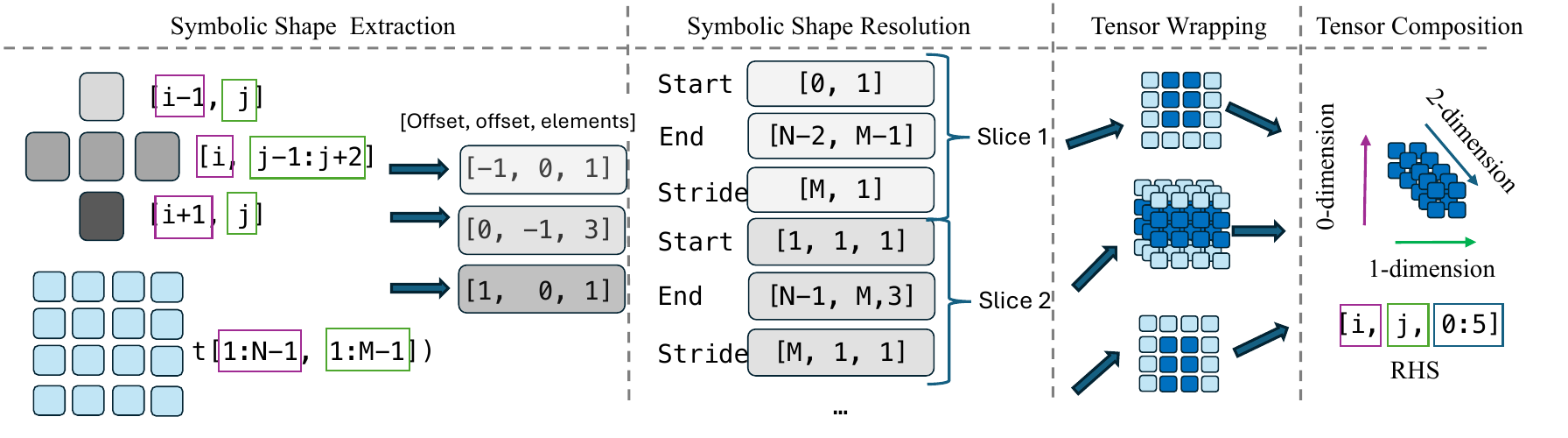}
   \caption{Steps of the \hpacml{} compiler runtime system performed during the data mapping of \texttt{ifnctr} (example code ~\ref{fig:hpacml_tensor_functor_example_1}).}
   \label{fig:hpac_impl}
   \vspace{-1em}
\end{figure*}

\hpacml{} implements the data bridge by closely co-designing the compiler
and runtime system. The runtime system implements 
custom data structures and the compiler performs a sequence of calls to runtime functions to declare, allocate, and fill those structures appropriately.

The parser, semantic analysis, and abstract syntax tree (AST) of Clang are extended to support the \verb|tensor| directive with its functor, map, and update clauses. These clauses require new expression types in the AST for multi-dimensional slices that support symbolic expressions. %
After parsing and semantic analysis of the \verb|functor| clause in the \verb|tensor| directive, code generation instantiates a tensor.
In code generation, a functor is effectively a structure representing views or slices on top of application memory locations. 
Until the functor is applied to memory, the pointer to application memory is undefined and symbolic expressions remain abstract.
As such, code generation generates two descriptors to be used by the runtime library one for the LHS and one the RHS of the functor. Each represents an \verb|ss-specifier| (Figure \ref{lst:syntax}).

The descriptors are a collection of slices. For example,  \verb|([i-1, j],| \verb|[i+1, j],| \verb|[i, j-1:j+2])| contains three slices. 
Every slice describes a view of memory through some pointer that will be provided when the functor is mapped to a memory location. A structure describes a slice and contains fields for the number of dimensions, shape, and strides. 
Symbolic constants (\verb|s-constant|), such as \verb|i, j|, are encoded in an AST node during parsing and semantic analysis to be assigned a unique identifier. The \hpacml{} code generator refers to them to compute possible ranges of values for these symbolic constants. The compiler bundles the LHS and RHS expressions and associates them with the \verb|functor-id| provided by the clause.

The tensor map clause creates a view of some user-specified application memory. Supporting the \verb|map| clause requires a sequence of steps to determine the base pointer, shape, and strides for converting user memory to a tensor.
Figure~\ref{fig:hpac_impl} illustrates these steps for the example in Figure~\ref{fig:hpacml_tensor_functor_example_1}. 
We demonstrate the process for \verb|map(ifnctr(t[1:N-1, 1:M-1]))|, where \verb|ifnctr| corresponds to the LHS \verb|[i, j, 0:5]| and RHS \verb|([i-1, j], [i+1, j], [i, j-1:j+2])| 

During \emph{Symbolic Shape Extraction}, the code generator extracts a \emph{shape} vector that contains one offset per dimension to apply to the base pointer, and the number of elements to retrieve. 
Figure~\ref{fig:hpacml_tensor_functor_example_1} has three slice descriptors, each referring to $2$-dimensional data.
For every slice, code generation associates the functor dimensions with the dimensions of the provided ranges in the \verb|map| clause. 
It then computes the offset of the first element in the slice relative to the first element in the base pointer, taking into account the minimum value of the range. In this example, this corresponds to \verb|t[1][1]|. 
Thus the slice \verb|[i-1][j]| will access the memory at location \verb|[0][1]| and the computed offset is $(-1, 0)$. Further, it computes the number of elements accessed by the slice.

Next follows \emph{Symbolic Shape Resolution}, which computes the shape 
of the resulting tensor when a slice is applied to memory. %
It computes the start index, end index, and stride for every dimension. Symbolic shape resolution adds a dimension of size 1 to slices with one element; for slices with more than one element (e.g., the slice \verb|[i,j-1:j+2]| has three elements), symbolic shape resolution adds a new dimension sized with the number of elements.

Following, the code generation moves to \emph{Tensor Wrapping}, in which it combines the per-slice 
descriptors computed during \emph{symbolic shape resolution} with the offsets computed during \emph{symbolic shape extraction}. It adds the offsets to each dimension's start and end ranges. %
This step creates a data structure compatible with most tensor libraries (Torch, TensorFlow, NumPy), as every slice now describes a view of some memory region.
So far no memory copies are needed: code generation creates lightweight wrappers around existing memory. %
Next, the \hpacml{} backend materializes these slices as library-specific tensors. %
Each RHS slice has been applied to application data, yielding one tensor for each.

These RHS tensors are combined in the \emph{Tensor Composition} step to create a single LHS tensor.
If more than one dimension was added in the symbolic shape resolution, they are flattened to create a tensor whose shape matches the LHS. %
Then, the RHS tensors are concatenated, and the mapping is complete.

The runtime system handles mapping in both directions between the application and tensor memory spaces. When the tensor map clause uses the \verb|from| specifier, the runtime creates a mapping from the tensor memory space to the application memory space. That involves the symbolic shape extraction, resolution, and tensor wrapping steps, which create a collection of tensors wrapping user data, without requiring the tensor reshaping and composition steps.%

\subsection{Execution Control}

\hpacml{} extends HPAC's \verb|approx| directive 
parsing and semantic analysis to recognize the 
\verb|ml|, \verb|model|, and \verb|database|
clauses.
HPAC captures the associated code statement following the \verb|approx| directive 
and outlines it in a new function. The wrapped region is replaced by a call to the HPAC runtime system and the compiler forwards a function pointer to the outlined function to the runtime system.

\hpacml{} extends HPAC's runtime API entry point to accept two file paths given as C-strings,
pointing to files. The first file, given by the \verb|database| clause, points
to a storage location to use during data collection, while the second file
refers to a model given in the \verb|model| clause.
Code generation for the \verb|ml| clause checks the \verb|ml-mode| identifier. When the identifier is set to \verb|infer|, code generation instructs the runtime to perform model inference throughout the application run. 
When \verb|ml-mode| is set to \verb|collect|, the compiler instructs the runtime to collect data throughout the application execution. 
For the \verb|predicated| identifier, the compiler generates code to evaluate a boolean expression passed as a second optional argument to the \verb|ml| clause. When the argument expression evaluates to \verb|false|, the runtime performs data collection; when it is set to \verb|true| the runtime performs inference.

At execution time, the runtime system can access the accurate execution path by invoking 
the function outlined by the compiler. \hpacml{} uses this during data collection.
Briefly, the runtime system maps the memory regions annotated as inputs with the \verb|in| clause to tensors using the data bridge
and stores them into an HDF5 file~\cite{HDF5}. Once the inputs are stored,
the outlined region is invoked. When it finishes, the outputs provided to the \verb|out| clause are mapped from memory and stored in the file.
The output HDF5 file contains an independent HDF5 group for every annotated region. Within each group \hpacml{} stores the following datasets:
\begin{inparaenum}
    \item inputs gathered by the data bridge;
    \item outputs gathered by the data bridge; and 
    \item execution time of the wrapped code region.
\end{inparaenum}

Storing both the data and the execution time of the outlined region in the HDF5 file
supports ML development workflows in the architecture search/model training phase of supervised machine learning. Developers can load the database using user-friendly HDF5 interfaces, such as \verb|h5py| in Python. They can then use the stored inputs and outputs to create training/validation datasets for model training. The included runtime information can be used to assess performance-accuracy trade-offs without executing the application. An ML engineer uses the database and associated runtime information to train a model with sufficient runtime performance and accuracy offline. The trained model is saved to disk in a language-agnostic format such as TorchScript for online inference in the application.

When the application instructs the \hpacml{} runtime system to perform inference,
the backend loads the model file if it has not already been loaded.
The implementation uses Torch, the C++ API of PyTorch, to load models. 
Once the model is loaded, \hpacml{} uses the data bridge to transform data 
from the application space to the tensor space and calls the model inference engine. The inference output tensor is converted back to the application layout through the data bridge.

\section{Evaluation}

\begin{table}[t]
\caption{The benchmarks used to evaluate \hpacml{}.}
\scriptsize
\begin{tabular}{p{0.15\columnwidth} p{0.75\columnwidth}}
\toprule
\textbf{Benchmark} & \textbf{Description}   \\
\midrule
MiniBUDE~\cite{Poenaru2021:Performance}   &  Executes virtual screening in molecular docking, assessing poses to predict ligand-protein binding energy using an empirical forcefield.  {\bfseries QOI:} Ligand-protein binding energy for each pose. {\bfseries Metric:} MAPE \\ \midrule
Binomial Options~\cite{Podlozhnyuk2008:Binomial}   & Iteratively calculates the price for a portfolio of American stock options at multiple time points before expiration.  {\bfseries QOI:} The computed prices. {\bfseries Metric:} RMSE\\ \midrule
Bonds~\cite{Grauer-Gray2013:Accelerating}             & Calculates bond valuations and interest payments for fixed-rate bonds with a flat forward curve. {\bfseries QOI:} The accrued interest for each bond.  
{\bfseries Metric:} RMSE
\\ \midrule
MiniWeather~\cite{Norman2020:MiniWeather}   & Simulates atmospheric dynamics through essential weather and climate modeling equations, emphasizing buoyant force impacts. {\bfseries QOI:} Simulation state variables (density, $x$ momentum, $y$ momentum, and potential temperature) at each gridpoint. 
{\bfseries Metric:} RMSE
\\ \midrule
ParticleFilter~\cite{Rodinia}             & Statistical estimation of a target object's location given noisy measurements. {\bfseries QOI:} The location of the object. 
{\bfseries Metric:} RMSE
\\ \bottomrule
\end{tabular}
\label{tab:benchmarks}
\end{table}

\hpacml{} provides an easy-to-use approach to collecting training data and deploying a model within an application.
In this section, we evaluate the complexity required to annotate the benchmarks, the one-time overhead of data collection, and finally, the performance overheads induced by our runtime implementation to invoke
the inference engine.

Table~\ref{tab:benchmarks} lists the evaluated CUDA benchmarks and the Quantity of Interest (QoI) for each.
The evaluation platform is equipped with one $64$-core AMD Epyc 7763 CPU and $4\times$ NVIDIA A100 GPUs, each with $40$GB of memory and $108$ SMs.
We use CUDA 11.8, PyTorch 2.2.0, and compile all benchmarks with the \hpacml{} compiler. 

\subsection{Code Complexity of Using \hpacml{}}

For every benchmark we aggressively annotate application regions with \hpacml{} 
to maximize potential performance gains. We use Lines of Code (LoC) as a metric of the complexity
of using \hpacml{}. 
We use \verb|clang-format| with the LLVM style to
format the sources and report the additional Lines of Code
required by \hpacml{}.
Table~\ref{tab:loc_overhead} details the total LoCs for every benchmark and LoC required to use \hpacml{} along with the number of directives.

\begin{table}[]
\caption{Application source code impact of \hpacml{}.}
\label{tab:loc_overhead}
\resizebox{\columnwidth}{!}{%
\begin{tabular}{llll}
\toprule
                 & \multicolumn{1}{c}{Total LoC} & \hpacml{} LoC & \hpacml{} Directives\\ \midrule
MiniBUDE         & 539                               & 8                  & 4                \\
Binomial Options & 455                                & 8                   & 4                \\
Bonds            & 2248                              & 9                 & 4                \\
MiniWeather      & 905                               & 3                 & 3                \\
ParticleFilter   & 1025                               & 5                  & 4                \\ \bottomrule
\end{tabular}%
}
\vspace{-1em}
\end{table}

Overall, we observe an average LoC increase of less than $2\%$ across all benchmarks.
MiniBUDE, Binomial Options, Bonds, and ParticleFilter each take 4 directives: input/output tensor functor declarations, a tensor mapping for the input, and the \verb|approx ml| directive. 
MiniWeather is an iterative solver using the same memory for an iteration's input/output, so we use the \verb|inout| clause.  %
In summary, a developer should expect to add up to 
two directives for the inputs and outputs to 
declare the tensor functor and the mapping between data layouts. An additional directive is required
to annotate the code to be replaced by the model. Using the same source code, \hpacml{} can 
perform data collection and model inference. 
In cases of identical memory layouts across application
 variables, functors can be re-used, further reducing code complexity.

\subsection{Data Collection}

\begin{table}[]
\centering
\caption{Data collection overhead.}
\label{tab:capture_overhead}
\resizebox{\columnwidth}{!}{%
\begin{tabular}{llll}

\toprule
                 & \multicolumn{1}{c}{Original Runtime} & Runtime With Data Collection & Collected Data Size (MB)  \\ \midrule
MiniBUDE         & 23.15s                               & 23.25s                  & 29                \\
Binomial Options & 2.02s                                & 2.21s                   & 268               \\
Bonds            & 0.96s                              & 1.18s                 & 303                \\
MiniWeather      & 0.65ms                               & 28.92ms                 & 19               \\
ParticleFilter   & 3.43 ms                               & 3.51 ms                  & 852                \\ \bottomrule
\end{tabular}%
}
\end{table}

Following best practices~\cite{Goodfellow2016:Deep}, we split the application datasets distributed with each benchmark into
$2$ different sets: a \emph{training/validation} set and a \emph{test} set.
The training/validation set is used for model training and hyperparameter tuning; the test set is used to evaluate the models when deployed with \hpacml{}.

We execute the \hpacml{} enhanced applications with the
\emph{training/validation} inputs and set the condition in the
\verb|ml(predicated:bool-expr)| clause to false, instructing the \hpacml{}
runtime system to collect data for future model training.
Collected data are stored in HDF5 files, a suitable format for ML
engineers to consume.
Effectively, inner dimensions correspond to the application's input-output
tensors and the outer dimension corresponds to a unique data identifier of this
set in the data collection ensemble.
The collected data are directly readable by the built-in PyTorch data loaders.

Table~\ref{tab:capture_overhead} details the overhead of capturing data for model training.
This overhead ranges between $1.01\times$ and $44.6\times$ over a normal application run.
Typically, this overhead is amortized, as ML experts reuse those data
when searching for a suitable model. %
This overhead is worthwhile after training such a model and significantly accelerating the application.

\subsection{Nested Bayesian-driven Model Exploration}

\hpacml{} facilitates easy use and deployment of different surrogate models. 
Additionally, the trade-offs between application accuracy and performance are
application/model-specific.
The model's architecture (i.e., how layers and nodes are structured and
connected) and training hyperparameters, such as the \emph{learning rate}, greatly affect the
the model's inference speed, prediction accuracy, and overall fidelity.
Exploring all possible model architectures and hyperparameters is practically
impossible, since they form an intractable, combinatorial space.
Hence, we devise an automated, guided approach to generate efficient models, 
leveraging application domain expertise to sub-select candidate ML architectures and Bayesian Optimization for searching architecture and training (hyper-)parameters.

We define a neural architecture space for each
benchmark to search for well-performing models.
We confine the model architecture based on our application domain expertise, e.g. we use convolutional layers to capture spatial information in the video frames of ParticleFilter.
Table~\ref{tab:searchspace} details each benchmark's neural architecture search space, while
Table~\ref{tab:searchspace_hypers} defines the space of feasible hyperparameter values.

\begin{table}[]
\caption{Search space used for neural architecture search
The \emph{Feature Multiplier} parameter denotes a factor that reduces the neuron count across hidden layers. 
}
\label{tab:searchspace}
\resizebox{\columnwidth}{!}{%
\begin{tabular}{@{}ll|ll@{}}
\toprule
\multicolumn{2}{l|}{\textbf{MiniBUDE}}                        & \multicolumn{2}{l}{\textbf{Binomial Options}, \textbf{Bonds}}                     \\ \midrule
\textit{Num. Hidden Layers}         & $[2, 12]$              & \textit{Hidden 1 Features}     & $[5, 512]$        \\
\textit{Hidden 1 Size}            & $64, 128,\ldots, 4096$                   & \textit{Hidden 2 Features} & $[0, 512]$ \\
\textit{Feature Multiplier}           & $[0.1, 0.8]$ & & \\ \midrule
\multicolumn{2}{l|}{\textbf{MiniWeather}}                      & \multicolumn{2}{l}{\textbf{ParticleFilter}}                     \\ \midrule
\textit{Conv. Layer 1 Kernel Size}        & $[2, 8]$  & \textit{Conv. Kernel Size; Conv. Stride}   & $[2,14]$             \\
\textit{Conv. Layer 1 Output Channels} & $[4,8]$ & \textit{Maxpool Kernel Size}                   & $[1, 10]$      \\
\textit{Conv. Layer 2 Kernel Size} & $[0,6]$ & \textit{FC 2 Size}                  & $[0, 128]$ \\
\end{tabular}%
}
\vskip 1em
\caption{Search space used for BO hyperparameter tuning.}
\label{tab:searchspace_hypers}
\resizebox{\columnwidth}{!}{%
\begin{tabular}{ll|ll}
\hline
\multicolumn{2}{l|}{\emph{Learning Rate}: $[1\times10^{-4}, 1\times10^{-2}]$} & \multicolumn{2}{l}{\emph{Weight Decay}: $[1\times10^{-4}, 1\times10^{-1}]$} \\ \hline
\multicolumn{2}{l|}{\emph{Dropout}: $[0, 0.8]$}             & \multicolumn{2}{l}{\emph{Batch Size}: $[32, 512]$}       \\ \hline
\end{tabular}%
}
\end{table}

Briefly, we employ
Bayesian Optimization (BO)~\cite{Mockus1989:Bayesian} to identify efficient model architectures and respective hyperparameters. BO streamlines model selection by using Gaussian Processes to predict the performance of an objective function, focusing on areas with potential improvement. Through an iterative process involving an acquisition function, BO efficiently explores and exploits the search space to find optimal solutions. This method is particularly effective for tasks like neural network tuning, where evaluations are costly
~\cite{Wu2019:Hyperparameter,Kandasamy2018:Neural}.

We use a nested, two-level, multi-objective BO for neural architecture search
and hyperparameter tuning. 
The search process looks for a neural architecture and its hyperparameters that jointly minimize inference latency \emph{and} validation error.
The outer level proposes neural architectures, while the inner level tunes hyperparameters to minimize validation error for the architecture.
The inner level produces hyperparameters that minimize validation error; the model architecture determines inference speed.

\begin{figure}
\centering
   \includegraphics[width=0.9\linewidth, trim={0 2em 0 0},clip]{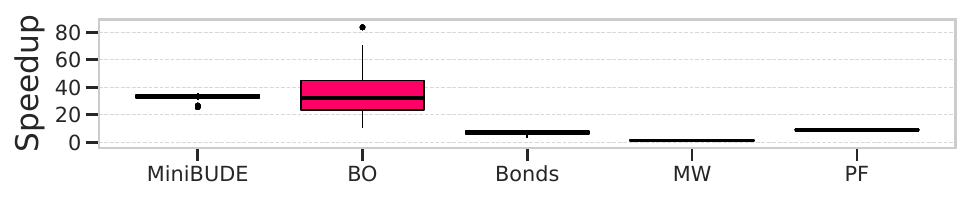}
   \includegraphics[width=0.9\linewidth]{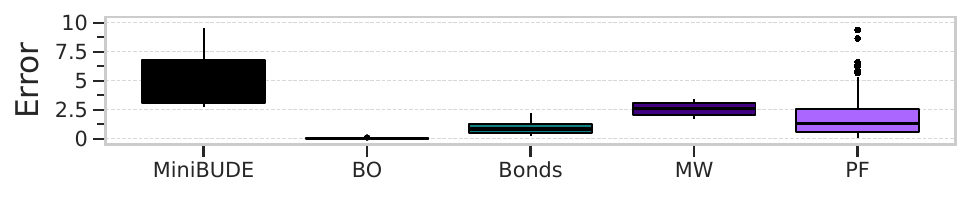}
   \caption{End-to-end application speedup and error of \hpacml{} enhanced applications.}
   \label{fig:summary}
   \vspace{-1em}
\end{figure}

The outer-level search is executed for $100$ iterations, with
early stopping when it doesn't find a faster or more accurate model for five consecutive trials.
The inner level tunes training hyperparameters for $30$ iterations.
Faithful to good practice, we train/evaluate all models \emph{only} using
the \emph{training/validation} data from the data collection phase.

We implement this approach as a workflow using  Parsl~\cite{Babuji2019:Parsl}
and employ the Adaptive Experimentation platform~\cite{Bakshy2018:Adaptive} for 
the Bayesian search.
The search convergence is fast, typically earlier than 16 trials.
Overall, our automated method explored $5130$ models: MiniBUDE $990$, Binomial
Options $900$, Bonds $540$, MiniWeather $780$, and ParticleFilter $1920$.

\subsection{Performance Benefits from \hpacml{} Model Deployment}
\label{ssec:performance_eval}

After model search and training, we integrate \hpacml{} using the lowest validation error models generated by the nested BO search (cutoff is error $<$ 10) and evaluate the application's error as observed at the final computed QoI.
MiniBUDE's error metric is MAPE; all other benchmarks use RMSE.
 Modifications in the application source code are minor: one sets the \verb|cond| in the
\verb|ml(infer:<cond-expr>)| to \verb|True| and adds the \verb|model(string-expr)| clause specifying the path of the generated model.

We evaluate \hpacml{} on two metrics: end-to-end speedup over the original GPU-accelerated application and quality loss. We run each configuration 20 times and use the (arithmetic) mean runtime, yielding statistically significant results.
Speedup is relative to end-to-end application runtime and includes all required data transfers and transformations.
Quality loss is computed using the QoI for each
application and measuring how much the DNN surrogate changes this QoI.

Figure~\ref{fig:summary} shows results of speedup and validation error across the lowest validation error models for all programs.
Overall, deploying models with \hpacml{} substantially decreases application execution time. 
We observe up to $83.6\times$ speedup, with a geometric mean of maximum possible speedup of $13.0\times$. 
Regarding validation error, (see Figure~\ref{fig:summary}-top), BO finds
several models below the cutoff of error $<$ 10.
Yet, whether a given error results in acceptable QoI is
application-specific, as discussed in the next section that explores per-program results.

\begin{figure}
   \centering
   \includegraphics[width=0.85\linewidth]{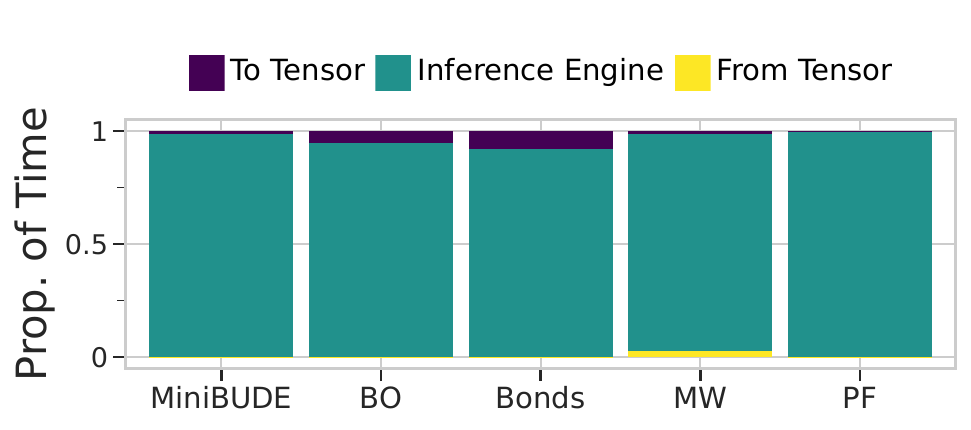}
   \caption{Proportion of time required for each of the primary operations performed by \hpacml{} in inference mode: mapping between tensors and memory and performing inference.}
   \label{fig:runtime_overheads}
\end{figure}

Further, Figure~\ref{fig:runtime_overheads} shows a breakdown of execution time within
the \hpacml{} runtime library, divided by time spent mapping tensors (to/from memory) and
inference time with the fastest model, which is delegated to the ML backend.
The overhead of \hpacml{} is between $0.01\%$ and $8\%$, compared to
the latency of the inference engine, which depends solely on the evaluation of the model by the ML backend.

\section{Observations}
\label{sec:evaluation}

We discuss the speedup and QoI accuracy results using the \hpacml{} programming model to extend programs for surrogate model execution
and the extensive ML model search campaign it made possible. We center the discussion around \emph{four observations} that summarize key insights.

\begin{figure}
   \centering
   \includegraphics[width=0.65\linewidth]{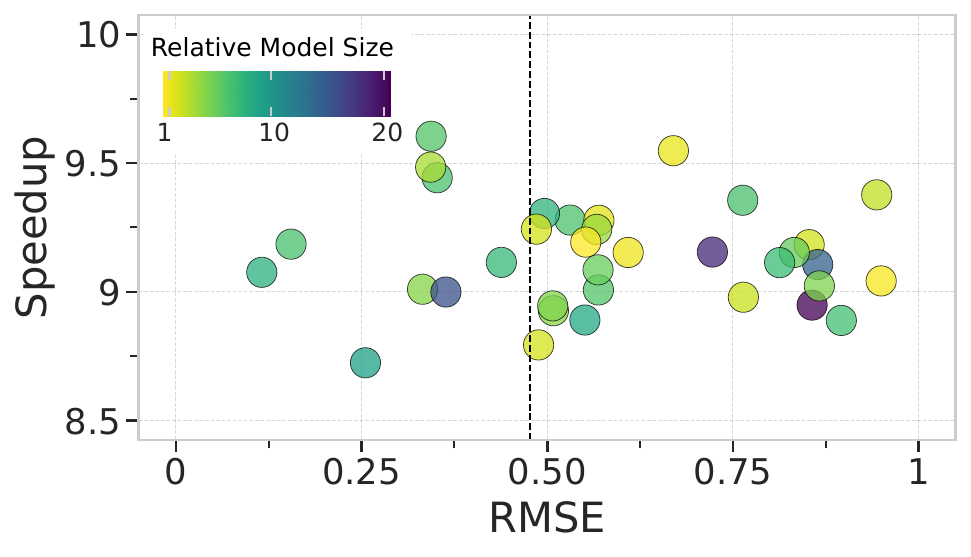}
\caption{ParticleFilter results for the models with RMSE $\le 1$. The black vertical line marks the RMSE of the original particle filter algorithm. Points are colored according to their number of parameters relative to the model with the fewest parameters.}%
\label{fig:particlefilter_results}
\vspace{-1em}
\end{figure}

\emph{Observation 1: An ML Model can outperform custom algorithmic approximation in \textbf{both execution time and accuracy}.}
Particlefilter (PF) approximates the location of a
target object, given noisy measurements of that target’s location and an estimate of
the object’s path using a Bayesian framework.
The application outputs the ground-truth object location and the particle filter's estimate.
The \hpacml{} version of PF captures the ground-truth values to create the training dataset.

Our BO search algorithm finds several models that accelerate the original application, as shown in Figure~\ref{fig:particlefilter_results}.
Briefly, NN surrogate modeling speeds up execution between $8.67\times$ and $9.60\times$ (end-to-end) over the original application implementing the algorithmic approximation.
Additionally, surrogates can reduce the output's RMSE to as low as $0.12$, whereas the original approximation's RMSE is $0.5$.
The surrogate model replaces the entire ParticleFilter algorithm, which includes three distinct GPU kernels, with a CNN that estimates the object's location in each video frame. 
CNN inference is up to $23.3\times$ faster than the kernels it
replaces.
While end-to-end speedup is subject to Amdahl's law, surrogate modeling speeds up the application by an order of magnitude.
As such, ParticleFilter is an
excellent application for replacing algorithmic approximation with surrogate
modeling approximation and aligns well with the known strengths of CNNs for image/video tasks.

\begin{figure*}[t]
    \centering
    \begin{subfigure}[c]{0.3\linewidth}
        \includegraphics[width=\linewidth]{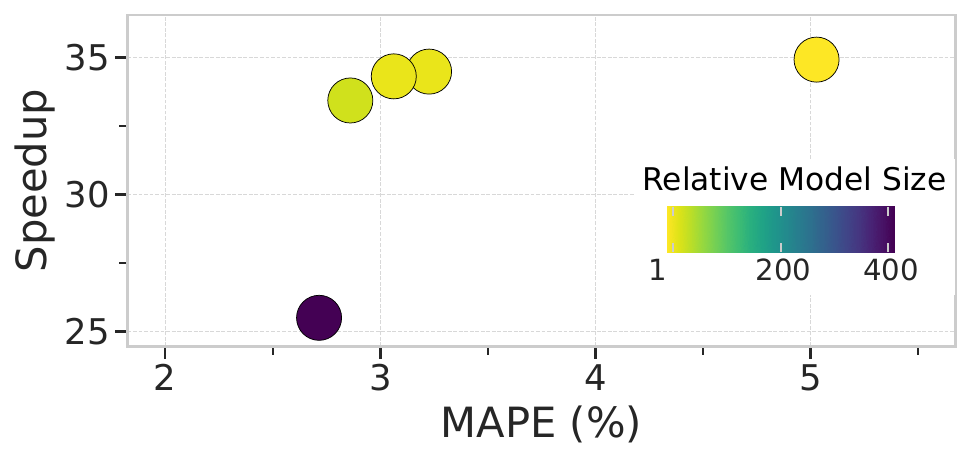}
        \caption{MiniBUDE}
        \label{fig:minibude_results}
    \end{subfigure}
    \begin{subfigure}[c]{0.3\linewidth}
        \includegraphics[width=\linewidth]{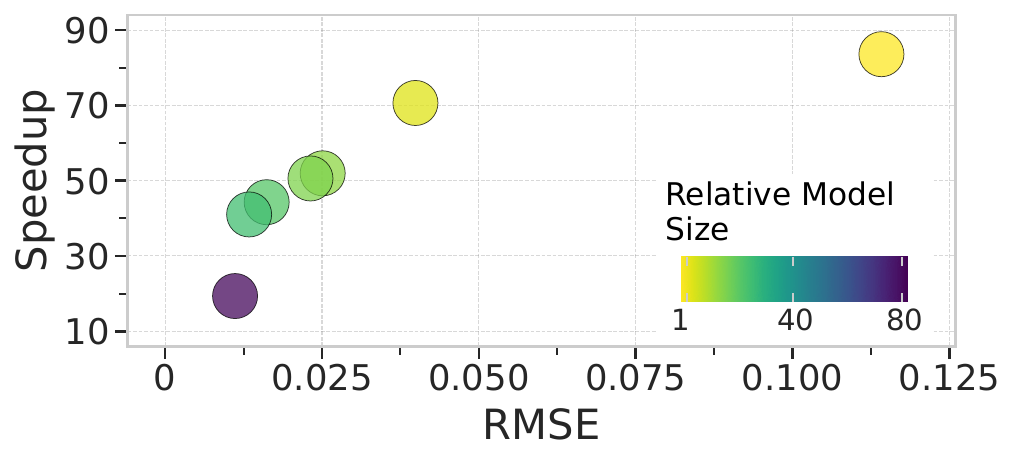}
        \caption{Binomial Options}
        \label{fig:bo_results}
    \end{subfigure}
    \begin{subfigure}[c]{0.3\linewidth}
        \includegraphics[width=\linewidth]{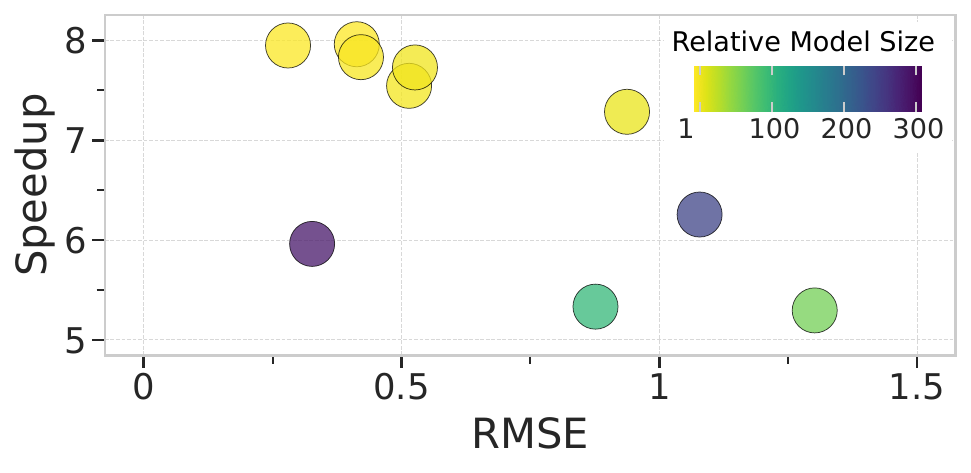}
        \caption{Bonds}
        \label{fig:bonds_results}
    \end{subfigure}
    \caption{Speedup vs Accuracy. Points are colored according to the number of parameters relative to the model with the fewest.}
    \label{fig:rsbench_amd}
    \vspace{-1.6em}
\end{figure*}

\begin{figure*}
    \centering
    \begin{subfigure}{0.3\linewidth}
    \centering
    \vspace{1em}
    \includegraphics[scale=0.38]{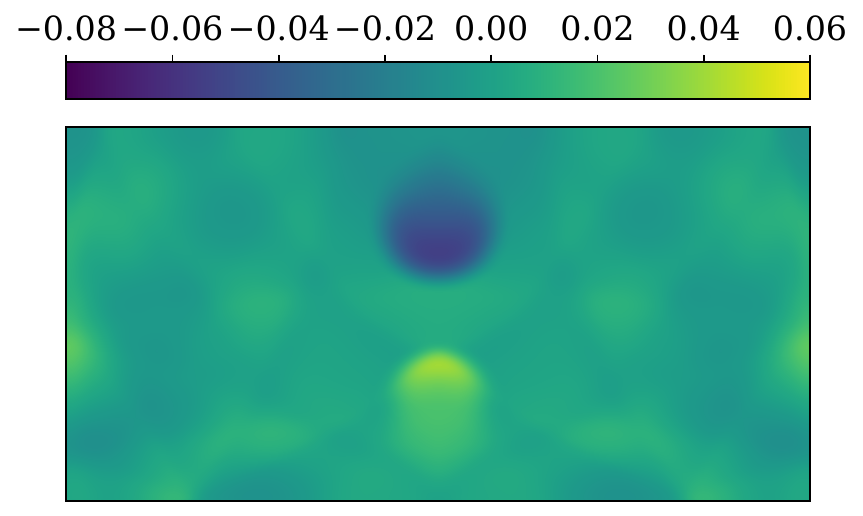}
    \caption{Original Simulation}
    \label{fig:mw_orig}
    \end{subfigure}
    \begin{subfigure}{0.3\linewidth}
    \centering
    \includegraphics[scale=0.38]{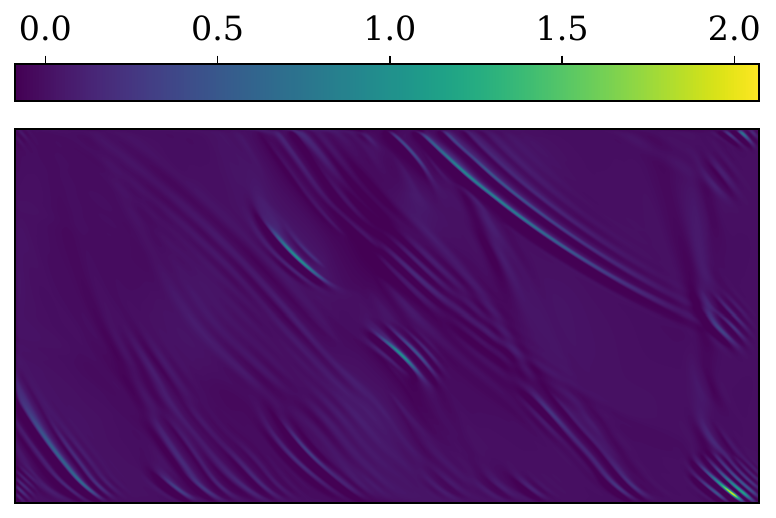}
    \caption{Only model inference simulation}
    \label{fig:mw_entire}
    \end{subfigure}
    \begin{subfigure}{0.3\linewidth}
    \centering
    \vspace{1em}
    \includegraphics[scale=0.38]{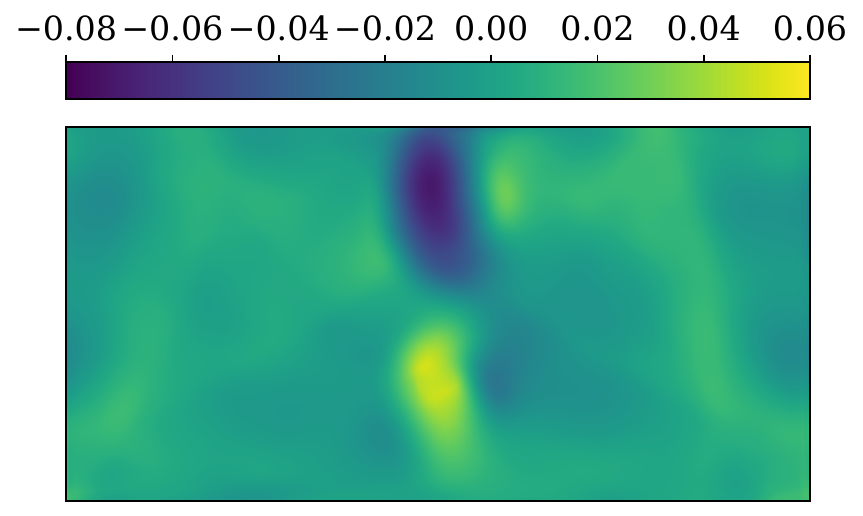}
    \caption{Mixed simulation.}
    \label{fig:mw_half}
    \end{subfigure}
    \begin{subfigure}{0.3\linewidth}
        \centering
        \includegraphics[width=\linewidth]{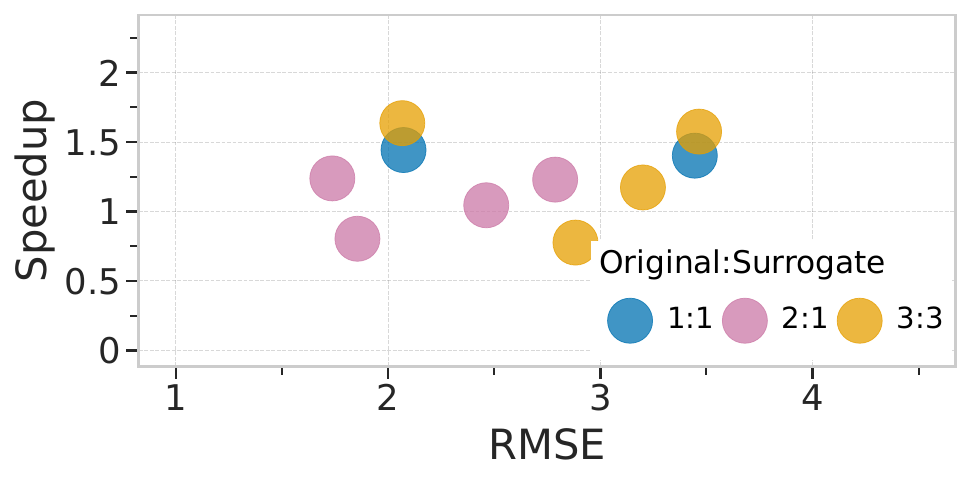}
        \caption{RMSE vs Speedup at timestep $1200$}
        \label{fig:lavamd_taf_amd}
    \end{subfigure}
    \begin{subfigure}{0.3\linewidth}
        \centering
        \includegraphics[width=\linewidth]{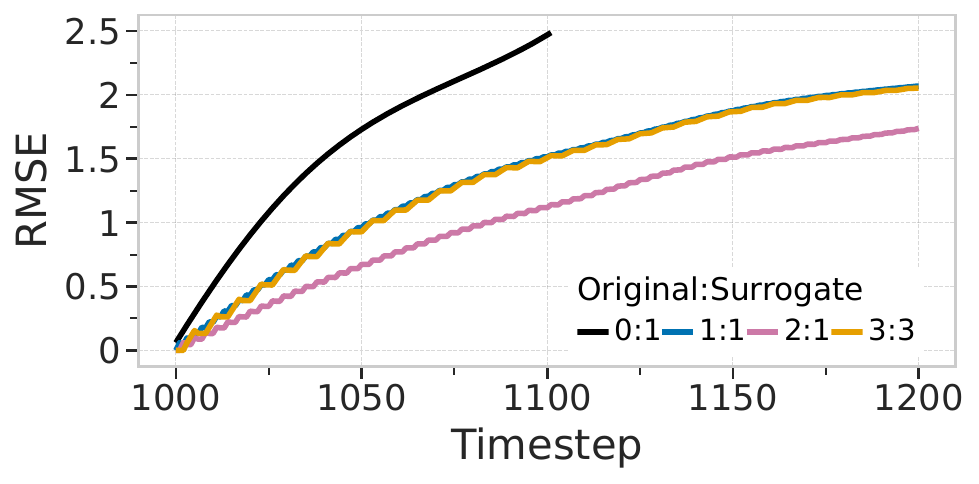}
        \caption{Per-timestep RMSE}
        \label{fig:lavamd_iact_amd}
    \end{subfigure}
         \begin{subfigure}{0.3\linewidth}
    \centering
    \includegraphics[width=\linewidth]{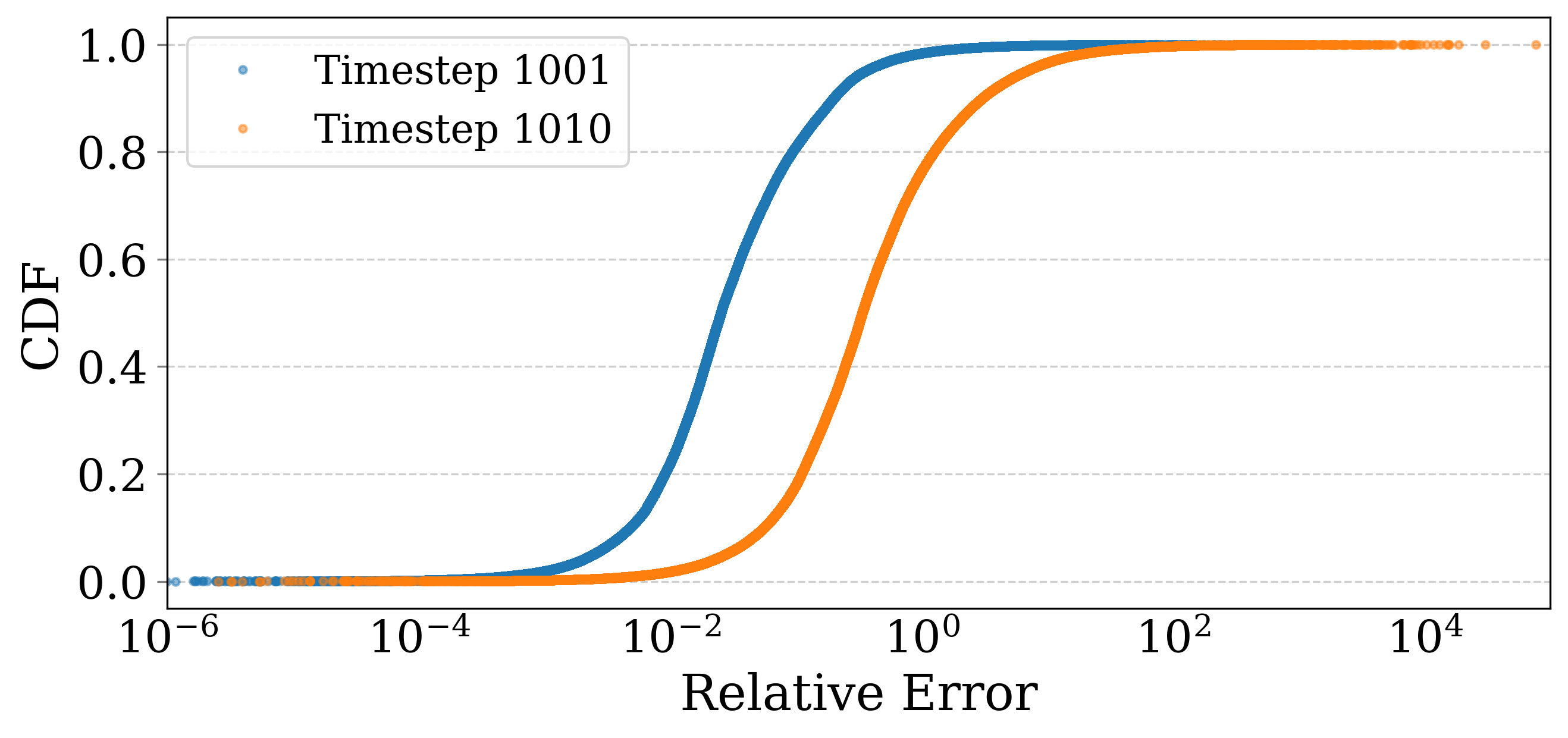}
        \caption{CDF of relative error}
        \label{fig:mw_cdf}
    \end{subfigure}
    \caption{
    Top shows MiniWeather simulation outputs at timestep 1200.
    All plots use the original solution until timestep 1000 because models are trained on the first 1000. In (b), we replace every timestep with model inference, in (c) we interleave model inference with the original timestep every other step.
    (d) Speedup vs. RMSE: This panel compares speedup against RMSE across various Original:Surrogate configurations, where `Original' denotes simulation iterations performed with the original timestep evaluation, and the next `Surrogate' loops are executed using the surrogate model. (e) Timestep vs RMSE for different configurations. (f) CDF of relative error from the surrogate model predictions at timestep 1001 versus error at timestep 1010.}
    \label{fig:interleave}
\end{figure*}

\emph{Observation 2:  Surrogate models can improve performance by increasing hardware utilization.}
While this motivates the use of surrogate models, we quantify it more precisely using MiniBUDE.
MiniBUDE predicts complex molecular structures and estimates their binding affinities. 
Those binding affinities are estimated by rotating and moving a ligand into different positions, called poses, and estimating the interaction strength between a ligand in each pose and a target protein.
Its original implementation is a compute-bound molecular docking mini-app whose kernel achieves $33.5\%$ of the device's peak compute throughput and $6.13\%$ of peak memory bandwidth.

By contrast, the fastest model achieves $47.2 \%$ of peak compute throughput and $31.5 \%$ of peak memory bandwidth, demonstrating much more efficient use of hardware resources.
 MiniBUDE accesses memory with a scattered pattern, whereas NN
inference is implemented by dense linear algebra operations, for which there are
highly optimized vendor software libraries that execute near-optimally on GPU
architectures.
Hence, execution of the surrogate model extracts higher performance both by
better alignment with the hardware architecture features and by reducing
computational load.

To quantify the effects of model size on accuracy and speedup,
Figure~\ref{fig:minibude_results} shows the Mean Absolute Percentage Error
(MAPE) of QoI and speedup for all Pareto-optimal
models.
The largest, slowest model achieves $25.5\times$ end-to-end speedup with
$2.71\%$ MAPE, while the fastest model worsens MAPE to $6.82\%$ MAPE but
increases speedup to $35\times$. 
While this result of MiniBUDE is expected -- larger models reduce speed up but are
more accurate -- this is not always the case, as our next observation details.

\emph{Observation 3: Automation complements human ML domain expertise for
exploring the trade-offs of model architecture, fidelity of the computed QoI,
model generalization, and overfitting.}
Binomial Options (Figure~\ref{fig:bo_results}) exhibits the same trade-offs as MiniBUDE: larger models are slower and more accurate.
However, the range of application speedup is different.
The smallest model speeds Binomial Options up by $83.59\times$, with an RMSE of $0.114$. 
Conversely, the largest model is much slower, with a speedup of $19.36\times$
(4.3$\times$ reduction) but with an RMSE of $0.0111$, an order of magnitude
improvement over the fastest one.
Quantifying this trade-off is greatly facilitated by the automation brought forward by \hpacml{} for applying surrogate modeling to enable this campaign.

Contrasting MiniBUDE and Binomial Options, larger models do not always
increase accuracy for Bonds, as Figure~\ref{fig:bonds_results} shows.
The fastest model is the most accurate, yielding an RMSE of $0.280$ and
$7.96\times$ speedup, surpassing the largest model (second-best for accuracy)
with an RMSE of $0.320$ and a speedup of $6.00\times$.
Generally, when larger models are less accurate than their smaller counterparts, it suggests that overfitting to the training data has occurred.
\hpacml{} makes it easier to identify and mitigate such cases.
The standard way to reduce overfitting is to augment the dataset, either by collecting more data or increasing the complexity of the dataset.
\hpacml{} makes data augmentation easy through its built-in data collection mechanisms.

\emph{Observation 4: In iterative auto-regressive scenarios, the model's error propagates across iterations and disturbs the application's QoI. \hpacml{} provides primitives to interleave model executions with accurate evaluations, thereby decreasing error.}
MiniWeather serves as a representative benchmark for atmospheric dynamics simulation programs, a domain where NN surrogate modeling has been particularly challenging~\cite{meena2023surrogate}.
In this context, even small inaccuracies introduced by the surrogate approximation can rapidly accumulate, leading to highly unstable and unrealistic simulations.
Our findings, despite the use of BO search to identify high-fidelity models, align with and support these previously reported difficulties.
In more detail, Figures~\ref{fig:mw_orig} and \ref{fig:mw_entire} present the visual output 
of MiniWeather after timestep $1200$ of the original simulation and of a simulation
using the model with the highest fidelity for all timesteps starting from timestep $1000$.
Although the surrogate model approximation adds little error in most grid locations, this error propagates and amplifies in later iterations.
Figure~\ref{fig:mw_cdf} shows the CDF of relative error in timestep 1001, after the first iteration using surrogate modeling, and timestep 1010.
At timestep 1001, the RMSE is 0.0003. Most error comes from few domain locations: $80\%$ of domain locations have relative error less than $0.09$ while $90\%$ have a relative error less than $0.2$.
\emph{However, using the model for $10$ consecutive times steps shifts the error distribution right by nearly an order of magnitude.} 
At timestep, 1010, $80\%$ of domain locations have a relative error of less than 1.25 while $90\%$ have a relative error of less than 3.04. 

\hpacml{} can be used to interleave model evaluations
with the original analytical algorithm on different timesteps.
Specifying the \texttt{if} clause controls usage of the surrogate model, enabling it in a user-specified fraction of the simulation's timesteps. 
Figures~\ref{fig:mw_half}~through~\ref{fig:mw_cdf} show results interleaving accurate time steps with surrogate inference in \hpacml{} to reveal performance and accuracy trade-offs for several interleaving configurations.
Interleaving model approximations with original outputs significantly decreases error, at the expense of performance improvement from surrogate modeling.
Nonetheless, this class of computation is especially hard to approximate with surrogates.
Figure~\ref{fig:mw_half} depicts the output of interleaving one original application timestep with one model timestep. Although the visual result is closer to the accurate one, the observed difference is still high.

\section{Conclusion}\label{sec:conclusion}
\hpacml{} is a novel pragma-based programming model for embedding ML surrogates in scientific applications.
The design and implementation of \hpacml{} and its key abstractions yield a high-productivity, high-performance method to integrate ML surrogate models into scientific applications with a few lines of code. Using \hpacml{}, we perform a comprehensive evaluation of $5$ GPU-accelerated HPC benchmarks testing thousands of models, demonstrating the versatility of \hpacml{} and exploring the trade-offs between model architecture, accuracy, and performance.
\hpacml{} bridges the gap between ML infrastructure and scientific applications, enabling researchers to rapidly evaluate ML surrogates in applications. This accelerated experimentation facilitates fundamental research on the interactions between ML and scientific computing. Application developers who want to improve application performance in the post-Moore era will find \hpacml{} a valuable tool in their belt.

\section*{Acknowledgements}
 The authors would like to thank the anonymous reviewers for
 their valuable comments and helpful suggestions. 

The views and opinions of the authors do not necessarily reflect those of the U.S. government or Lawrence Livermore National Security, LLC neither of whom nor any of their employees make any endorsements, express or implied warranties or representations or assume any legal liability or responsibility for the accuracy, completeness, or usefulness of the information contained herein. This work was prepared by LLNL under Contract DE-AC52-07NA27344 (LLNL-CONF-862478) and was supported by both the LLNL LDRD Program under Project No. 22-SI-004. This material is based in part upon work supported by the Department of Energy, National Nuclear Security Administration, under Award Number DE-NA0003963. This work used the Delta system at the National Center for Supercomputing Applications through allocation ASC050025 from the Advanced Cyberinfrastructure Coordination Ecosystem: Services \& Support (ACCESS) program, which is supported by National Science Foundation grants \#2138259, \#2138286, \#2138307, \#2137603, and \#2138296.
\bibliographystyle{IEEEtran}
\bibliography{ref.bib}

% Generated by IEEEtran.bst, version: 1.14 (2015/08/26)
\begin{thebibliography}{10}
\providecommand{\url}[1]{#1}
\csname url@samestyle\endcsname
\providecommand{\newblock}{\relax}
\providecommand{\bibinfo}[2]{#2}
\providecommand{\BIBentrySTDinterwordspacing}{\spaceskip=0pt\relax}
\providecommand{\BIBentryALTinterwordstretchfactor}{4}
\providecommand{\BIBentryALTinterwordspacing}{\spaceskip=\fontdimen2\font plus
\BIBentryALTinterwordstretchfactor\fontdimen3\font minus
  \fontdimen4\font\relax}
\providecommand{\BIBforeignlanguage}[2]{{%
\expandafter\ifx\csname l@#1\endcsname\relax
\typeout{** WARNING: IEEEtran.bst: No hyphenation pattern has been}%
\typeout{** loaded for the language `#1'. Using the pattern for}%
\typeout{** the default language instead.}%
\else
\language=\csname l@#1\endcsname
\fi
#2}}
\providecommand{\BIBdecl}{\relax}
\BIBdecl

\bibitem{Galantai2000:Theory}
A.~Galántai, ``The theory of newton's method,'' \emph{Journal of Computational
  and Applied Mathematics}, vol. 124, no.~1, 2000, numerical Analysis 2000.
  Vol. IV: Optimization and Nonlinear Equations.

\bibitem{Funaro2008:Polynomial}
D.~Funaro, \emph{Polynomial approximation of differential equations}.\hskip 1em
  plus 0.5em minus 0.4em\relax Springer Science \& Business Media, 2008,
  vol.~8.

\bibitem{Sidiroglou2011:Managing}
S.~Sidiroglou-Douskos, S.~Misailovic, H.~Hoffmann, and M.~Rinard, ``Managing
  performance vs. accuracy trade-offs with loop perforation,'' in
  \emph{Proceedings of the 19th ACM SIGSOFT symposium and the 13th European
  conference on Foundations of software engineering}, 2011.

\bibitem{Hoffmann2009:Using}
H.~Hoffmann, S.~Misailovic, S.~Sidiroglou, A.~Agarwal, and M.~Rinard, ``Using
  code perforation to improve performance, reduce energy consumption, and
  respond to failures,'' 2009.

\bibitem{Menon2018:ADAPT}
H.~{Menon}, M.~O. {Lam}, D.~{Osei-Kuffuor}, M.~{Schordan}, S.~{Lloyd},
  K.~{Mohror}, and J.~{Hittinger}, ``{ADAPT}: Algorithmic differentiation
  applied to floating-point precision tuning,'' in \emph{SC18: International
  Conference for High Performance Computing, Networking, Storage and Analysis},
  2018, pp. 614--626.

\bibitem{Laguna2019:GPUMixer}
I.~Laguna, P.~C. Wood, R.~Singh, and S.~Bagchi, ``Gpumixer: Performance-driven
  floating-point tuning for gpu scientific applications,'' in
  \emph{International Conference on High Performance Computing}, 2019.

\bibitem{mishra2014iact}
A.~K. Mishra, R.~Barik, and S.~Paul, ``{iACT}: A software-hardware framework
  for understanding the scope of approximate computing,'' in \emph{Workshop on
  Approximate Computing Across the System Stack (WACAS)}, 2014.

\bibitem{tziantzioulis2018temporal}
G.~Tziantzioulis, N.~Hardavellas, and S.~Campanoni, ``Temporal approximate
  function memoization,'' \emph{IEEE Micro}, vol.~38, no.~4, 2018.

\bibitem{pal2019using}
A.~Pal, S.~Mahajan, and M.~R. Norman, ``Using deep neural networks as
  cost-effective surrogate models for super-parameterized e3sm radiative
  transfer,'' \emph{Geophysical Research Letters}, vol.~46, no.~11, pp.
  6069--6079, 2019.

\bibitem{kochkov2021machine}
D.~Kochkov, J.~A. Smith, A.~Alieva, Q.~Wang, M.~P. Brenner, and S.~Hoyer,
  ``Machine learning--accelerated computational fluid dynamics,''
  \emph{Proceedings of the National Academy of Sciences}, vol. 118, no.~21, p.
  e2101784118, 2021.

\bibitem{parasyris2021hpac}
K.~Parasyris, G.~Georgakoudis, H.~Menon, J.~Diffenderfer, I.~Laguna,
  D.~Osei-Kuffuor, and M.~Schordan, ``{HPAC}: evaluating approximate computing
  techniques on hpc openmp applications,'' in \emph{Proceedings of the
  International Conference for High Performance Computing, Networking, Storage
  and Analysis}, 2021, pp. 1--14.

\bibitem{fink2023:hpacoffload}
\BIBentryALTinterwordspacing
Z.~Fink, K.~Parasyris, G.~Georgakoudis, and H.~Menon, ``Hpac-offload:
  Accelerating hpc applications with portable approximate computing on the
  gpu,'' in \emph{Proceedings of the International Conference for High
  Performance Computing, Networking, Storage and Analysis}, ser. SC '23.\hskip
  1em plus 0.5em minus 0.4em\relax New York, NY, USA: Association for Computing
  Machinery, 2023. [Online]. Available:
  \url{https://doi.org/10.1145/3581784.3607095}
\BIBentrySTDinterwordspacing

\bibitem{froehlich2018towards}
S.~Froehlich, D.~Gro{\ss}e, and R.~Drechsler, ``Towards reversed approximate
  hardware design,'' in \emph{2018 21st Euromicro Conference on Digital System
  Design (DSD)}, 2018.

\bibitem{Rehman2016}
S.~Rehman, W.~{El-Harouni}, M.~Shafique, A.~Kumar, J.~Henkel, and J.~Henkel,
  ``Architectural-space exploration of approximate multipliers,'' in \emph{2016
  {{IEEE}}/{{ACM International Conference}} on {{Computer}}-{{Aided Design}}
  ({{ICCAD}})}, 2016.

\bibitem{kulkarni2011trading}
P.~Kulkarni, P.~Gupta, and M.~Ercegovac, ``Trading accuracy for power with an
  underdesigned multiplier architecture,'' in \emph{2011 24th Internatioal
  Conference on VLSI Design}, 2011.

\bibitem{kahng2012accuracy}
A.~B. Kahng and S.~Kang, ``Accuracy-configurable adder for approximate
  arithmetic designs,'' in \emph{Proceedings of the 49th annual design
  automation conference}, 2012.

\bibitem{gupta2011impact}
V.~Gupta, D.~Mohapatra, S.~P. Park, A.~Raghunathan, and K.~Roy, ``Impact:
  Imprecise adders for low-power approximate computing,'' in \emph{IEEE/ACM
  International Symposium on Low Power Electronics and Design}, 2011.

\bibitem{chippa2014scalable}
V.~K. Chippa, D.~Mohapatra, K.~Roy, S.~T. Chakradhar, and A.~Raghunathan,
  ``Scalable effort hardware design,'' \emph{IEEE Transactions on Very Large
  Scale Integration (VLSI) Systems}, vol.~22, no.~9, 2014.

\bibitem{esmaeilzadeh2012architecture}
H.~Esmaeilzadeh, A.~Sampson, L.~Ceze, and D.~Burger, ``Architecture support for
  disciplined approximate programming,'' in \emph{Proceedings of the
  seventeenth international conference on Architectural Support for Programming
  Languages and Operating Systems}, 2012.

\bibitem{hedge1999energy}
R.~Hegde and N.~Shanbhag, ``Energy-efficient signal processing via algorithmic
  noise-tolerance,'' in \emph{1999 International Symposium on Low Power
  Electronics and Design}, 1999.

\bibitem{sampson2011enerj}
A.~Sampson, W.~Dietl, E.~Fortuna, D.~Gnanapragasam, L.~Ceze, and D.~Grossman,
  ``Enerj: Approximate data types for safe and general low-power computation,''
  \emph{ACM SIGPLAN Notices}, vol.~46, no.~6, pp. 164--174, 2011.

\bibitem{parasyris2017significance}
K.~Parasyris, V.~Vassiliadis, C.~D. Antonopoulos, S.~Lalis, and N.~Bellas,
  ``Significance-aware program execution on unreliable hardware,'' \emph{ACM
  Transactions on Architecture and Code Optimization (TACO)}, vol.~14, no.~2,
  pp. 1--25, 2017.

\bibitem{Hoffmann2009}
H.~Hoffmann, S.~Misailovic, S.~Sidiroglou, M.~Rinard, and A.~Agarwal, ``Using
  {{Code Perforation}} to {{Improve Performance}}, {{Reduce Energy
  Consumption}}, and {{Respond}} to {{Failures}},'' Tech. Rep.
  MIT-CSAIL-TR-2009-042, 2009.

\bibitem{goiri2015approxhadoop}
I.~Goiri, R.~Bianchini, S.~Nagarakatte, and T.~D. Nguyen, ``Approxhadoop:
  Bringing approximations to mapreduce frameworks,'' in \emph{Proceedings of
  the Twentieth International Conference on Architectural Support for
  Programming Languages and Operating Systems (ASPLOS)}, 2015.

\bibitem{michie1968memo}
D.~Michie, ``“memo” functions and machine learning,'' \emph{Nature}, vol.
  218, 1968.

\bibitem{keramidas2015clumsy}
G.~Keramidas, C.~Kokkala, and I.~Stamoulis, ``Clumsy value cache: An
  approximate memoization technique for mobile gpu fragment shaders,'' in
  \emph{Workshop on approximate computing (WAPCO’15)}, 2015.

\bibitem{samadi2014paraprox}
M.~Samadi, D.~A. Jamshidi, J.~Lee, and S.~Mahlke, ``Paraprox: Pattern-based
  approximation for data parallel applications,'' in \emph{Proceedings of the
  19th international conference on Architectural support for programming
  languages and operating systems}, 2014.

\bibitem{vassiliadis2016towards}
V.~Vassiliadis, J.~Riehme, J.~Deussen, K.~Parasyris, C.~D. Antonopoulos,
  N.~Bellas, S.~Lalis, and U.~Naumann, ``Towards automatic significance
  analysis for approximate computing,'' in \emph{2016 IEEE/ACM International
  Symposium on Code Generation and Optimization (CGO)}, 2016, pp. 182--193.

\bibitem{vassiliadis2016exploiting}
V.~Vassiliadis, C.~Chalios, K.~Parasyris, C.~D. Antonopoulos, S.~Lalis,
  N.~Bellas, H.~Vandierendonck, and D.~S. Nikolopoulos, ``Exploiting
  significance of computations for energy-constrained approximate computing,''
  \emph{International Journal of Parallel Programming}, vol.~44, no.~5, pp.
  1078--1098, 2016.

\bibitem{vassiliadis2015programming}
V.~Vassiliadis, K.~Parasyris, C.~Chalios, C.~D. Antonopoulos, S.~Lalis,
  N.~Bellas, H.~Vandierendonck, and D.~S. Nikolopoulos, ``A programming model
  and runtime system for significance-aware energy-efficient computing,''
  \emph{ACM SIGPLAN Notices}, vol.~50, no.~8, pp. 275--276, 2015.

\bibitem{grigorian2015brainiac}
B.~Grigorian, N.~Farahpour, and G.~Reinman, ``{BRAINIAC}: Bringing reliable
  accuracy into neurally-implemented approximate computing,'' in \emph{2015
  IEEE 21st International Symposium on High Performance Computer Architecture
  (HPCA)}, 2015.

\bibitem{mcafee2015emeuro}
L.~McAfee and K.~Olukotun, ``Emeuro: A framework for generating multi-purpose
  accelerators via deep learning,'' in \emph{2015 IEEE/ACM International
  Symposium on Code Generation and Optimization (CGO)}.\hskip 1em plus 0.5em
  minus 0.4em\relax IEEE, 2015, pp. 125--135.

\bibitem{Moreau2015:SNNAP}
T.~Moreau, M.~Wyse, J.~Nelson, A.~Sampson, H.~Esmaeilzadeh, L.~Ceze, and
  M.~Oskin, ``Snnap: Approximate computing on programmable socs via neural
  acceleration,'' in \emph{2015 IEEE 21st International Symposium on High
  Performance Computer Architecture (HPCA)}.\hskip 1em plus 0.5em minus
  0.4em\relax IEEE, 2015, pp. 603--614.

\bibitem{esmaeilzadeh2012neural}
H.~Esmaeilzadeh, A.~Sampson, L.~Ceze, and D.~Burger, ``Neural acceleration for
  general-purpose approximate programs,'' in \emph{2012 45th Annual IEEE/ACM
  International Symposium on Microarchitecture}, 2012.

\bibitem{Ho2022:Tensorox}
N.-M. Ho and W.-F. Wong, ``Tensorox: {{Accelerating GPU Applications}} via
  {{Neural Approximation}} on {{Unused Tensor Cores}},'' \emph{IEEE
  Transactions on Parallel and Distributed Systems}, vol.~33, no.~2, pp.
  429--443, Feb. 2022.

\bibitem{sharif2019approxhpvm}
H.~Sharif, P.~Srivastava, M.~Huzaifa, M.~Kotsifakou, K.~Joshi, Y.~Sarita,
  N.~Zhao, V.~S. Adve, S.~Misailovic, and S.~V. Adve, ``Approxhpvm: a portable
  compiler ir for accuracy-aware optimizations.'' \emph{Proc. ACM Program.
  Lang.}, no. OOPSLA, 2019.

\bibitem{sampson2015accept}
A.~Sampson, A.~Baixo, B.~Ransford, T.~Moreau, J.~Yip, L.~Ceze, and M.~Oskin,
  ``Accept: A programmer-guided compiler framework for practical approximate
  computing,'' \emph{University of Washington Technical Report UW-CSE-15-01 1
  (2)}, 2015.

\bibitem{baek2010green}
W.~Baek and T.~M. Chilimbi, ``Green: A framework for supporting
  energy-conscious programming using controlled approximation,'' in
  \emph{Proceedings of the 31st ACM SIGPLAN Conference on Programming Language
  Design and Implementation}, 2010, pp. 198--209.

\bibitem{Parasyris2022:Approximate}
K.~Parasyris, J.~Diffenderfer, H.~Menon, I.~Laguna, J.~Vanover, R.~Vogt, and
  D.~Osei-Kuffuor, ``Approximate computing through the lens of uncertainty
  quantification,'' in \emph{SC22: International Conference for High
  Performance Computing, Networking, Storage and Analysis}, 2022.

\bibitem{Dong2023:AutoHPCNet}
\BIBentryALTinterwordspacing
W.~Dong, G.~Kestor, and D.~Li, ``Auto-hpcnet: An automatic framework to build
  neural network-based surrogate for high-performance computing applications,''
  in \emph{Proceedings of the 32nd International Symposium on High-Performance
  Parallel and Distributed Computing}, ser. HPDC '23.\hskip 1em plus 0.5em
  minus 0.4em\relax New York, NY, USA: Association for Computing Machinery,
  2023, p. 31–44. [Online]. Available:
  \url{https://doi.org/10.1145/3588195.3592985}
\BIBentrySTDinterwordspacing

\bibitem{Myers2024:Amrex}
A.~Myers, W.~Zhang, A.~Almgren, T.~Antoun, J.~Bell, A.~Huebl, and A.~Sinn,
  ``Amrex and pyamrex: Looking beyond ecp,'' 2024.

\bibitem{paszke2019pytorch}
A.~Paszke, S.~Gross, F.~Massa, A.~Lerer, J.~Bradbury, G.~Chanan, T.~Killeen,
  Z.~Lin, N.~Gimelshein, L.~Antiga \emph{et~al.}, ``Pytorch: An imperative
  style, high-performance deep learning library,'' \emph{Advances in neural
  information processing systems}, vol.~32, 2019.

\bibitem{abadi2016tensorflow}
M.~Abadi, A.~Agarwal, P.~Barham, E.~Brevdo, Z.~Chen, C.~Citro, G.~S. Corrado,
  A.~Davis, J.~Dean, M.~Devin \emph{et~al.}, ``Tensorflow: Large-scale machine
  learning on heterogeneous distributed systems,'' \emph{arXiv preprint
  arXiv:1603.04467}, 2016.

\bibitem{Bai2019:ONNX}
J.~Bai, F.~Lu, K.~Zhang \emph{et~al.}, ``Onnx: Open neural network exchange,''
  \url{https://github.com/onnx/onnx}, 2019.

\bibitem{VanEssen2015:LBANN}
\BIBentryALTinterwordspacing
B.~Van~Essen, H.~Kim, R.~Pearce, K.~Boakye, and B.~Chen, ``Lbann: livermore big
  artificial neural network hpc toolkit,'' in \emph{Proceedings of the Workshop
  on Machine Learning in High-Performance Computing Environments}, ser. MLHPC
  '15.\hskip 1em plus 0.5em minus 0.4em\relax New York, NY, USA: Association
  for Computing Machinery, 2015. [Online]. Available:
  \url{https://doi.org/10.1145/2834892.2834897}
\BIBentrySTDinterwordspacing

\bibitem{HDF5}
\BIBentryALTinterwordspacing
{The HDF Group}, ``{Hierarchical Data Format, version 5}.'' [Online].
  Available: \url{https://github.com/HDFGroup/hdf5}
\BIBentrySTDinterwordspacing

\bibitem{Poenaru2021:Performance}
A.~Poenaru, W.-C. Lin, and S.~McIntosh-Smith, ``A performance analysis of
  modern parallel programming models using a compute-bound application,'' in
  \emph{International Conference on High Performance Computing}.\hskip 1em plus
  0.5em minus 0.4em\relax Springer, 2021, pp. 332--350.

\bibitem{Podlozhnyuk2008:Binomial}
V.~Podlozhnyuk. (2007) Binomial option pricing model.

\bibitem{Grauer-Gray2013:Accelerating}
\BIBentryALTinterwordspacing
S.~Grauer-Gray, W.~Killian, R.~Searles, and J.~Cavazos, ``Accelerating
  financial applications on the gpu,'' in \emph{Proceedings of the 6th Workshop
  on General Purpose Processor Using Graphics Processing Units}, ser.
  GPGPU-6.\hskip 1em plus 0.5em minus 0.4em\relax New York, NY, USA:
  Association for Computing Machinery, 2013, p. 127–136. [Online]. Available:
  \url{https://doi.org/10.1145/2458523.2458536}
\BIBentrySTDinterwordspacing

\bibitem{Norman2020:MiniWeather}
\BIBentryALTinterwordspacing
M.~R. Norman and USDOE, ``miniweather,'' 3 2020. [Online]. Available:
  \url{https://www.osti.gov//servlets/purl/1631691}
\BIBentrySTDinterwordspacing

\bibitem{Rodinia}
S.~Che, M.~Boyer, J.~Meng, D.~Tarjan, J.~W. Sheaffer, S.-H. Lee, and
  K.~Skadron, ``Rodinia: A benchmark suite for heterogeneous computing,'' in
  \emph{2009 IEEE International Symposium on Workload Characterization
  (IISWC)}, 2009.

\bibitem{Goodfellow2016:Deep}
I.~Goodfellow, Y.~Bengio, and A.~Courville, \emph{Deep Learning}.\hskip 1em
  plus 0.5em minus 0.4em\relax MIT Press, 2016,
  \url{http://www.deeplearningbook.org}.

\bibitem{Mockus1989:Bayesian}
J.~Mockus, \emph{Bayesian Approach to Global Optimization: Theory and
  Applications}, ser. Mathematics and its Applications.\hskip 1em plus 0.5em
  minus 0.4em\relax Kluwer Academic Publishers, 1989, vol.~37.

\bibitem{Wu2019:Hyperparameter}
J.~Wu, X.-Y. Chen, H.~Zhang, L.-D. Xiong, H.~Lei, and S.-H. Deng,
  ``Hyperparameter optimization for machine learning models based on bayesian
  optimization,'' \emph{Journal of Electronic Science and Technology}, vol.~17,
  no.~1, pp. 26--40, 2019.

\bibitem{Kandasamy2018:Neural}
K.~Kandasamy, W.~Neiswanger, J.~Schneider, B.~Poczos, and E.~P. Xing, ``Neural
  architecture search with bayesian optimisation and optimal transport,''
  \emph{Advances in neural information processing systems}, vol.~31, 2018.

\bibitem{Babuji2019:Parsl}
Y.~Babuji, A.~Woodard, Z.~Li, D.~S. Katz, B.~Clifford, R.~Kumar, L.~Lacinski,
  R.~Chard, J.~Wozniak, I.~Foster, M.~Wilde, and K.~Chard, ``Parsl: Pervasive
  parallel programming in python,'' in \emph{28th ACM International Symposium
  on High-Performance Parallel and Distributed Computing (HPDC)}, 2019.

\bibitem{Bakshy2018:Adaptive}
E.~Bakshy, L.~Dworkin, B.~Karrer, K.~Kashin, B.~Letham, A.~Murthy, and
  S.~Singh, ``Ae: A domain-agnostic platform for adaptive experimentation,'' in
  \emph{NeurIPS Systems for ML Workshop}, 2018.

\bibitem{meena2023surrogate}
M.~G. Meena, M.~R. Norman, D.~M. Hall, and M.~S. Pritchard, ``Surrogate
  modeling of subgrid-scale effects in idealized atmospheric flows: A deep
  learned approach using high-resolution simulation data,'' \emph{Authorea
  Preprints}, 2023.

\end{thebibliography}

\ifincludeappendix
\def\hpacml{\textsc{hpac-ml}}

\twocolumn[%
{\begin{center}
\Huge
Appendix: Artifact Description/Artifact Evaluation        
\end{center}}
]

\appendixAD

\section{Overview of Contributions and Artifacts}

\subsection{Paper's Main Contributions}

\begin{itemize}
    \item[$C_1$] \hpacml{}, an annotation-based programming model for embedding ML surrogates in scientific applications. Application developers annotate parts of their application to replace them with a surrogate model. \hpacml{} enables developers to describe the data flow between application memory and ML data layouts using
    succinct syntax.
    \item[$C_2$] An implementation of the \hpacml{} programming model in Clang/LLVM v17 with a runtime system that uses Torch for inference and HDF5 for data collection.
    \item[$C_3$] A thorough evaluation on a representative suite of GPU-based HPC mini-apps and benchmarks. We find up to $83.6\times$ speedup while maintaining output quality with RMSE as low as $0.001$. 
    \item [$C_4$] \hpacml{} enables a large-scale exploration of the trade-offs between model size, inference speed, and output quality loss. To explore these trade-offs, we trained thousands of models.
\end{itemize}

\subsection{Computational Artifacts}
Our paper produces the following computational artifacts.

\begin{itemize}
    \item[$A_1$] Software environment for the \hpacml{} compiler/runtime system and experimental evaluation. \url{https://doi.org/10.5281/zenodo.13328344}
    \item[$A_2$] \hpacml{} compiler and runtime system implementation. \url{https://doi.org/10.5281/zenodo.12823916}
    \item[$A_3$] \hpacml{} benchmark programs. \url{https://doi.org/10.5281/zenodo.13328344}
    \item[$A_4$] \hpacml{} model training and evaluation infrastructure. \url{https://doi.org/10.5281/zenodo.13328369}
\end{itemize}

\begin{center}
\begin{tabular}{rll}
\toprule
Artifact ID  &  Contributions &  Related \\
             &  Supported     &  Paper Elements \\
\midrule
$A_1$   &  $C_1-C_4$ & Tables 1-3 \\
        &        & Figures 5-9\\
\midrule
$A_2$   &  $C_1-C_4$ & Tables 2-3 \\
        &        & Figures 5-9\\
\midrule
$A_3$   &  $C_3-C_4$ & Tables 1-3 \\
        &        & Figures 5-9\\
\midrule
$A_4$   &  $C_3-C_4$ & Figures 5-9 \\
\bottomrule
\end{tabular}
\end{center}

\section{Artifact Identification}

\newartifact
\artrel
This artifact contains the software environment used to build the compiler/runtime system and perform the experiments listed in the paper. Therefore, this artifact is a requirement for all the artifacts that follow.

\artexp 
After reproducing this artifact, the \hpacml{} compiler/runtime system can be built and evaluated.

\arttime
Setting up the artifact involves cloning the GitHub repository and navigating to the \verb|software_|\verb|env| directory. It is expected to take less than 1 minute. Use the included installation script to execute the artifact and install the software environment. This step will take up to 5 hours, depending on the number of cores used.

\artin

\artinpart{Hardware}
Our hardware evaluation platform has one $64$-core AMD Epyc 7763 CPU and $4\times$ NVIDIA A100 GPUs, each with $40$GB of memory and $108$ SMs.

Generally, the software environment can be built for NVIDIA GPUs and does not depend on the interconnect and other platform details.

\artinpart{Software}
This artifact requires git. Our platform uses git version 2.39.3. Additionally, either Docker or Apptainer are required for the container. We have used Docker v26.1.4 and Apptainer v1.3.1. The provided installation script installs the other required software into the container. All packages are installed using gcc 11.4.0.

\artinpart{Datasets / Inputs}
There are no datasets or inputs for this artifact.

\artinpart{Installation and Deployment}
This artifact is installed and activated with the included \verb|setup.sh| and \verb|run_container.sh| scripts. 

\artcomp
After installation, the software environment can be activated by running the container with the provided \verb|run_container.sh| script.

\artout
This artifact performs no data analysis.

\newartifact

\artrel
This artifact is the implementation of the \hpacml{} programming model and runtime system as described in the paper. Therefore, all contributions are derived from this implementation. In particular, $C_3$ and $C_4$ use this implementation to perform our computational experiments on a target platform.

\artexp 
Reproducing this artifact means building the compiler and runtime system in the provided container. We have also shared container images with the pre-built compiler; during installation, the user can decide whether to build the container image or download an existing image using the installation script. The \hpacml{} compiler can then be used to compile programs using the \hpacml{} programming model; the runtime system implements the runtime operations of the programming model.

\arttime
On our platform, building the compiler/runtime system from scratch using $25$ CPU cores takes about $30$ minutes.

\artin

\artinpart{Hardware}
Our hardware evaluation platform is equipped with one $64$-core AMD Epyc 7763 CPU and $4\times$ NVIDIA A100 GPUs, each with $40$GB of memory and $108$ SMs.

Generally, the software environment can be built for NVIDIA GPUs and does not depend on the interconnect and other platform details.

\artinpart{Software}
The \hpacml{} compiler/runtime system use the software environment described in $A_1$. All packages are built and installed in a container by the installation process. Some packages of particular importance to this artifact include:
\begin{enumerate}
    \item PyTorch 2.2.0 \url{https://github.com/pytorch/pytorch}
    \item HDF5 1.14.3 \url{https://github.com/HDFGroup/hdf5}
    \item CUDA 11.8.89
\end{enumerate}

A complete list of software packages can be found in the configuration files associated with $A_1$.

\artinpart{Datasets / Inputs}
The only input required is the number of cores to use for building the software.

\artinpart{Installation and Deployment}
To deploy the compiler so it can be used to build programs, run the included \verb|source| \verb|hpac_|\verb|env|\verb|.sh| to add the compiler and its associated runtime libraries to the system path. The container sources this script as part of the startup process: when using the container, the user need not manually deploy the compiler.

\artcomp
This artifact is installed by cloning the associated git repository and running the \verb|setup|\verb|.sh| script that builds the \hpacml{} compiler and runtime system.

\artout
The build process produces versions of the \verb|clang| and \verb|clang++| compilers equipped to ingest code written in the \hpacml{} programming model and generate code as described in the paper.

\newartifact

\artrel
This artifact consists of the suite of programs modified to use the \hpacml{} programming model. This includes the benchmark programs gathered from the original sources and modified to include the \hpacml{} annotations. Additional modifications are added as needed to support the analysis workflow. They are compiled with the \hpacml{} compiler and run under the \hpacml{} runtime system to generate the per-benchmark results shown in the paper.

\artexp 
The user can navigate to the directory associated with each benchmark to view its source code and compile the original/approximated version using the provided makefiles.

\arttime
Each benchmark takes less than 30 seconds to build on our platform. To run each with data collection enabled requires up to $1$ minute for any of the benchmarks.

\artin

\artinpart{Hardware}
Our hardware evaluation platform is equipped with one $64$-core AMD Epyc 7763 CPU and $4\times$ NVIDIA A100 GPUs, each with $40$GB of memory and $108$ SMs.

Generally, the software environment can be built for NVIDIA GPUs and does not depend on the interconnect and other platform details.

\artinpart{Software}
We use the \hpacml{} compiler to build each benchmark; thus, the build process for these benchmarks assumes the software environment created in $A_1$ is activated, and the compiler has been built.

\artinpart{Datasets / Inputs}
All benchmarks except for Binomial Options and MiniBUDE generate their own data sets at runtime. For Binomial Options, we use a data set with $16$ million unique options; for MiniBUDE, our dataset includes $16$ million poses. These datasets are available at \url{https://doi.org/10.5281/zenodo.12751779}.

\artinpart{Installation and Deployment}
Each benchmark is compiled with the \hpacml{} programming model using the makefiles found with each.

\artcomp
The benchmarks can be run on their own by compiling them with the provided makefiles and running the generated binary. $A_4$ implements a workflow that automates the end-to-end process of training models and deploying them in each benchmark.
To gather training data, each benchmark is annotated with the \hpacml{} programming model, and data collection mode is enabled. Then, the benchmark is run and the input/output data are captured into an HDF5 database for offline training. We make all training data available at \url{https://doi.org/10.5281/zenodo.12751779}.

\artout
This artifact does not generate data to be analyzed.

\newartifact

\artrel
This artifact performs the neural architecture search driven by nested Bayesian Optimization described in the paper. The trained models are then embedded in the application, and the performance and accuracy are reported. These performance/accuracy values are reported via the plots given in the paper.

\artexp 
This artifact produces the following: 
\begin{inparaenum}
    \item the model with the best hyperparameters for a given neural architecture explored during the search in TorchScript format;
    \item the performance and accuracy that result from embedding each trained model in the application.
\end{inparaenum}
With this artifact, it is possible to reproduce the figures listed in the table. The results should indicate substantial runtime speedup with low accuracy loss for the applications when the model is used.

\arttime
We make the trained models for all benchmarks available at \url{https://doi.org/10.5281/zenodo.12751779}. Thus, they can be easily embedded in the applications to assess performance/accuracy trade-offs, skipping the neural architecture search. To perform the full BO-based neural architecture search takes between $50$ and $400$ GPU hours for a given benchmark. Assessing the performance and accuracy trade-offs for a given model takes between 1 and 5 minutes.

\artin

\artinpart{Hardware}
Our hardware evaluation platform is equipped with one $64$-core AMD Epyc 7763 CPU and $4\times$ NVIDIA A100 GPUs, each with $40$GB of memory and $108$ SMs.

Generally, the software environment can be built for NVIDIA GPUs and does not depend on the interconnect and other platform details.

\artinpart{Software}
This artifact assumes the container environment is being used. All software packages used for this artifact are included in the software environment. We note two packages important for this step:
\begin{enumerate}
    \item Adaptive Experimentation Platform v0.3.1: Performs Bayesian Optimization. \url{https://ax.dev/}
    \item Parsl v2024.02.26: Orchestrates the workflow steps required for the nested BO-based neural architectures search. \url{https://parsl-project.org/}
\end{enumerate}

\artinpart{Datasets / Inputs}
The training data for all benchmarks are located at: \url{https://doi.org/10.5281/zenodo.12751779}. These datasets are collected by running each benchmark with its corresponding input dataset using \hpacml{} in data collection mode. Each data set is split into two parts: a training/validation set used for training models and hyperparameter tuning and a test set used to evaluate the trained models on unseen data. Each dataset is large enough to have sufficient model training data and enough testing data to saturate device computational resources.

\artinpart{Installation and Deployment}
The Spack environment introduced in $A_1$ should be activated before running the artifact.

\artcomp
This artifact (\url{https://doi.org/10.5281/zenodo.13328369}) has two steps: model training and embedding. 
The model training step is located within the \verb|model_training| directory; embedding is in the \verb|benchmark_evaluation| directory.

The provided software performs each step end-to-end. Given the provided architecture search space configuration file, the model training step automatically performs the neural architecture search and saves the best models to the specified output directory for each benchmark. Given trained models, the benchmark evaluation script embeds each model in the application, runs it with and without the NN model used in the application, captures the accurate and approximated output, and calculates end-to-end speedup and output loss. This script also captures ancillary timing information, such as time spent on data transfers, model inference, etc.

Each step generates a \verb|csv| file containing the results. In the model evaluation stage, each benchmark is run $20$ times, with each instance given a line in the \verb|csv| output file.

The trained models are available for download, allowing you to skip the model training step. Each directory has a \verb|commands|\verb|.sh| script that performs the corresponding step.

\artout
The information displayed in the plots is derived directly from the output \verb|csv| files generated by the model evaluation stage for each benchmark. Each benchmark is run for $20$ iterations: we drop the first two and use the final $18$ to calculate the arithmetic average values used for speedup: accuracy results are constant for a given model.

\newpage
\appendixAE

\arteval{1}
\artin
To begin, clone the repository \url{https://doi.org/10.5281/zenodo.13328344}. Then, use the provided \verb|setup|\verb|.sh| script to build or download or build a container image using either Docker or Apptainer. 

This script also downloads all data necessary for training, and evaluating surrogate models.
Upon completion, the installation creates a file \verb|run_|\verb|container|\verb|.sh| that activates the container environment.

\artexpl{
Provide instructions for installing and compiling libraries and code. 
Offer guidelines on deploying the code to resources.
}

\artcomp
The script \verb|setup.sh| installs/compiles all necessary libraries and code into a container.

\artexpl{
Describe the experiment workflow. 
If encapsulated within a workflow description or equivalent (such as a makefile or script), clearly outline the primary tasks and their interdependencies. Detail the main steps in the workflow. Merely instructing to “Run script.sh” is inadequate.
}

\artout
This artifact does not contain any output for analysis.

\artexpl{
\begin{itemize}
    \item Provide a description of the expected results and a methodology for evaluating these results. 
    \item Explain how the expected results from the experiment workflow correlate with the contributions stated in the article. 
    \item For example, if the article presents results in a figure, the artifact evaluation should also produce a similar figure, depicting the same generalizable outcome. Authors must focus on these aspects to reduce the time required for others to understand and verify an artifact.
\end{itemize}
}

\arteval{2}
\artin
This artifact is built in the setup for Artifact 1 and installed in the \verb|hpacml| directory. The container includes the \hpacml{} compiler and runtime system, and sets the modified \verb|clang| and \verb|clang++| executables to the \hpacml{} compiler.

\artcomp
This artifact is executed by compiling programs with the \verb|clang++| executable and using the \verb|-fapprox| flag to compile programs. Note that the artifact includes Makefiles and driver scripts that automatically build all benchmakr programs.
\artout
This artifact does not contain any output for analysis.

\arteval{3}
\artin
The \hpacml{} benchmark suite is installed to the \verb|benchmarks| directory during the setup for Artifact 1. Each benchmark has a \verb|Makefile| to compile the non-approximated version and a \verb|Makefile.|\verb|approx| to compile the approximated version.

\artcomp
These artifacts are executed in the experimental workflow of Artifact 4.

\artout
This artifact does not itself contain any output for analysis.

\arteval{4}
\artin
This artifact is installed in the container setup in Artifact 1.
\artcomp
This artifact has two components: model training and benchmark evaluation.
Model training performs the nested Bayesian Optimization process outlined in the paper to explore accuracy/performance trade-offs.
This artifact stores the created models in a directory for each benchmark.
Because training for each benchmark takes between $50$ and $400$ GPU hours, we have included the trained models in the artifact. 
These trained models are used for benchmark evaluation.

For each model created for every benchmark, the benchmark evaluation step does the following with a Python script:
\begin{itemize}
    \item Build the accurate version of the benchmark and run it, capturing the runtime and outputs.
    \item Build the approximate version of the benchmark and run it, capturing the runtime and outputs.
    \item Compute error and speedup and store the results to a CSV file.
    \item Plot the results to create the accuracy/performance trade-off figures shown in the paper.
\end{itemize}

\artout
The CSV files contain model speedup and error, along with the timing of different events in the \hpacml{} runtime system, such as translating between application memory and NN tensors.
The format is explained futher in the \verb|README| file that accompanies the evaluation.

These files are used to create accuracy/speedup scatterplots, like the ones shown in the paper. 
This artifact does this automatically, executing the entire workflow and outputting a plot.

\fi
\end{document}